\documentclass[conference,10pt]{IEEEtran}
\usepackage{amsfonts, bm}
\usepackage{amssymb}
\usepackage{color,graphicx}
\usepackage{cite}
\usepackage[cmex10]{amsmath}
\usepackage{amsopn}
\usepackage{multirow}
\usepackage{bigstrut}
\usepackage{microtype}
\usepackage{verbatim}
\usepackage{stfloats}

\usepackage[ruled]{algorithm2e}

\IEEEoverridecommandlockouts

\def\gap{1.0ex}
\abovedisplayskip\gap
\belowdisplayskip\gap
\abovedisplayshortskip\gap
\belowdisplayshortskip\gap

\graphicspath{{figs/}}

\allowdisplaybreaks

\newtheorem{proposition}{Proposition}

\newtheorem{corollary}{Corollary}

\newtheorem{lemma}{Lemma}

\begin{document}

\sloppy

\title{{Joint Beam Training and Data Transmission Design for Covert Millimeter-Wave Communication}}


\author{Jiayu Zhang, Min Li, Shihao Yan, Chunshan Liu, Xihan Chen, Minjian Zhao and Philip Whiting~\thanks{Jiayu Zhang, Min Li, Xihan Chen and Minjian Zhao are with College of Information Science and Electronic Engineering, Zhejiang University, Hangzhou, 310027, China (e-mail: \{11731056, min.li, chenxihan, mjzhao\}@zju.edu.cn).
\par Shihao Yan and Philip Whiting are with School of Engineering, Macquarie University, Sydney, NSW 2109, Australia (e-mail: \{shihao.yan, philip.whiting\}@mq.edu.au).
\par Chunshan Liu is with School of Communication Engineering, Hangzhou Dianzi University, Hangzhou, 310018, China (e-mail: chunshan.liu@hdu.edu.cn).}}
\maketitle

\begin{abstract}
Covert communication prevents legitimate transmission from being detected by a warden while maintaining certain covert rate at the intended user. Prior works have considered the design of covert communication over conventional low-frequency bands, but few works so far have explored the higher-frequency millimeter-wave (mmWave) spectrum. The directional nature of mmWave communication makes it attractive for covert transmission. However, how to establish such directional link in a covert manner in the first place remains as a significant challenge. In this paper, we consider a covert mmWave communication system, where legitimate parties Alice and Bob adopt beam training approach for directional link establishment. Accounting for the training overhead, we develop a new design framework that jointly optimizes beam training duration, training power and data transmission power to maximize the effective throughput of Alice-Bob link while ensuring the covertness constraint at warden Willie is met. We further propose a dual-decomposition successive convex approximation algorithm to solve the problem efficiently. Numerical studies demonstrate interesting tradeoff among the key design parameters considered and also the necessity of joint design of beam training and data transmission for covert mmWave communication.
\end{abstract}

\begin{IEEEkeywords}
Beam alignment, beam training, covert communication, millimeter-wave communications, training-throughput tradeoff.
\end{IEEEkeywords}

\section{Introduction}
\par Covert communication~\cite{bash2015hiding}, also known as low-probability-of-detection (LPD) communication~\cite{BashGT13}, has emerged as a new security paradigm in wireless systems. Different from conventional physical-layer security methods, covert communication aims to hide the very existence of legitimate transmission from an adversary while maintaining a certain covert rate at the intended user. It therefore achieves a stronger security and privacy level, which is highly desired in the emerging 5G/IoT systems and advanced military networks~\cite{liu2018covert,yan2019low}.

\par Consider a classic covert communication setup, where Alice wishes to send message to Bob over a wireless channel, while ensuring that the probability of the transmission being detected by a warden Willie is small (i.e., a covertness constraint at Willie). For an additive white Gaussian noise (AWGN) channel, Bash \textit{et al.}~\cite{BashGT13} established that Alice can only send ${\mathcal O}(\sqrt{n})$ covert bits to Bob over $n$ channel uses. This square-root law has later been shown to also hold for a binary symmetric channel~\cite{CheBJ13} and a broader class of discrete memoryless channels~\cite{WangWZ16}. However, this square-root result can be further improved, for instance, by the use of an additional jammer to facilitate Alice's transmission~\cite{SobersBGTG17}. Note such scaling-law results are obtained for sufficiently large $n$.

\par Yan {\textit{et al.}}~\cite{yan2018delay} instead considered a delay-intolerant setup and studied the impact of finite $n$ on the covert communication performance. The optimality of Gaussian signalling has been then analyzed in delay-intolerant covert communications~\cite{yan2019gaussian}. Various other practical constraints, such as channel uncertainty~\cite{ShahzadZY17,Xu19} and noise uncertainty~\cite{he2017covert} have also been modeled and investigated in covert communications. In addition, more complicated scenarios that involve artificial noise~\cite{soltani2018covert}, multi-antenna nodes~\cite{LeeBMF14,AbdelazizK17,zheng2019multi,hu2019optimal,shahzad2019covert}, full-duplex nodes~\cite{shahzad2018achieving} and relay-assisted transmission~\cite{arumugam2018covert2,wang2018covert,hu2019covert} have also been considered.

\par The above works focused on the design of covert transmission over conventional low-frequency bands. Compared to these frequency bands, millimeter-wave band has much more under-utilized spectrum and has now been put forward as an important means to expand the capacity for mobile communications~\cite{xiao2017millimeter,liu2018Mag}. Due to its unfavorable propagation characteristics (such as high path loss and limited scattering), mmWave communication would heavily rely on beamforming transmission to ensure reliable links. The directional nature of the communication link makes it inherently suitable for covert transmission, because it is much more challenging for an adversary to overhear all of the communication.

\par While the potential of mmWave covert communication is conceivable, fundamental understanding and design guidelines for such system are still lacking. Related studies are scant except~\cite{cotton2009millimeter,Jamali19}, to the best of our knowledge. Reference~\cite{cotton2009millimeter} introduced a conceptual framework of mmWave soldier-to-soldier covert communications and discussed a few challenges at the physical layer and medium access control layer. However, it neither considered beamforming design that is crucial for the system nor provided rigorous quantification of the covertness level the framework can fundamentally achieve. Reference~\cite{Jamali19} considered a covert mmWave communication system, where Alice deploys dual independent antenna arrays, with one to form a beam towards Bob for covert data transmission and the other to form another beam towards Willie for jamming transmission. The outage probability and optimal covert rate of Alice-Bob link were characterized. However, this work only focused on the data transmission phase and did not address how the Alice-Bob directional link is established in the first place and whether or not the establishment of such link would require additional communication that might be detected by Willie.

\begin{figure*}[t]
	\centering
	\includegraphics[width=0.75\textwidth]{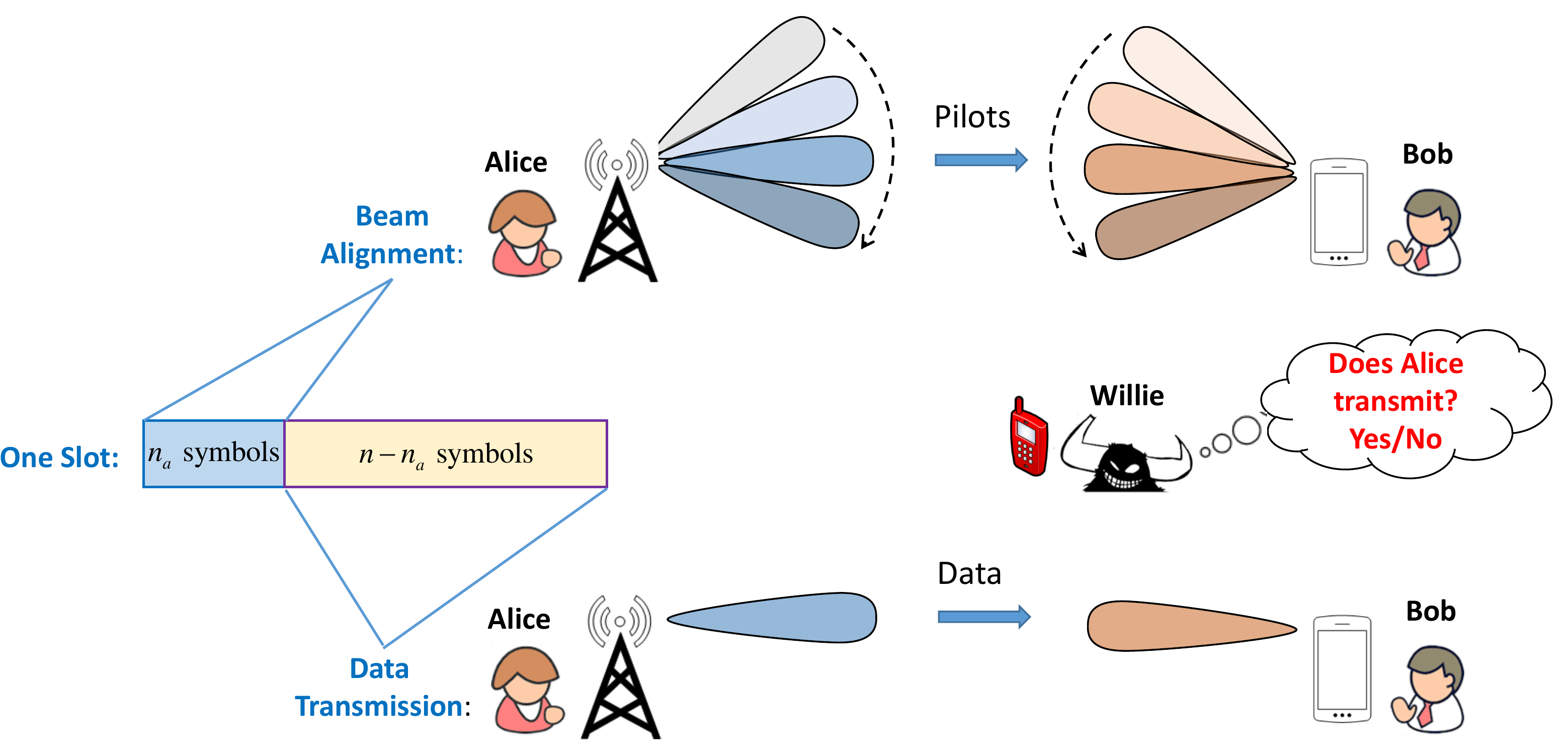}\caption{An illustration of the system model considered: Alice and Bob first carry out beam training to find the best beam pair that aligns with the dominant path of the channel, and then use this beam pair found for data transmission. This process is subject to the surveillance of warden Willie, who wishes to detect whether Alice transmits anything or not.}\label{fig:sysmodel}
\end{figure*}

\par Motivated by these observations, in this work, we consider a joint design of link establishment (beam alignment) and data transmission for covert mmWave communication. Specifically, we assume that both Alice and Bob are equipped with antenna arrays, while Willie is equipped with an omni-directional antenna to monitor all possible directions. Within a channel coherence time, Alice and Bob first take a commonly-used beam training approach~\cite{liu2017Jsac,li2019explore} to determine the best transmit-receive beam pair that aligns well with the channel, and then use this beam-pair found for subsequent data transmission. For a fixed coherence time, there is a tradeoff between the beam training duration and the effective throughput of the Alice-Bob link, since increasing the training duration can improve the beam alignment performance but at the expense of reducing time for data transmission. In addition, having larger training power and data transmission power can also contribute to an improvement of the throughput, however, this will increase Willie's chance to successfully detect the presence of the Alice-Bob communication. Hence, a fundamental question is: \textit{How to jointly optimize the beam training duration, training power and data transmission power to maximize the throughput of Alice-Bob link while ensuring the covertness constraint imposed on Willie is met? }

\par In this work, we address this question by assuming that Alice-Bob link is a Line-of-Sight (LOS) single-path channel for analytical tractability. With generalized flat-top beam codebooks and for exhaustive-search beam training, we derive a lower bound on the successful alignment probability as a function of beam training duration and training power. Based on this, we then develop a lower bound on the effective throughput of Alice-Bob link and study the training-throughput tradeoff optimization, subject to a covertness constraint at Willie. The resultant problem is highly nonconvex. To efficiently solve this problem, we exploit its structural properties and propose a \textit{Dual-decomposition Successive Convex Approximation} (DSCA) algorithm. Numerical results demonstrate an interesting tradeoff among the key design parameters considered and also the necessity of joint beam training and data transmission design for covert mmWave communication. The resultant optimal effective throughput for Alice-Bob link crucially depends on the covertness level targeted by the system.

\par The remainder of the paper is organized as follows. Section~\ref{sec:system:model} describes the mmWave covert communication model considered. Section~\ref{sec:joint:optimization} first characterizes the beam alignment and throughput performance of Alice-Bob link and the detection performance at Willie, and then moves on to study the optimized covert communication design. Numerical results are provided in Section~\ref{sec:numerical:results}, while conclusions are drawn in Section~\ref{sec:conclusions}.

\par \textit{Notation}: $(\cdot)^T$ denotes the matrix transpose, while $(\cdot)^\dag$ denotes the conjugate transpose. Given a vector $\mathbf x$, $\mathbf{x}[i]$ denotes the $ i $-th element of $\mathbf x$. For integers $z_1 \le z_2$, $\left[z_1:z_2\right]$ denotes the discrete interval $\{z_1,z_1+1,\cdots,z_2\}$. ${\mathcal{CN}}(0,\sigma^2)$ is a zero-mean complex Gaussian variable with variance $\sigma^2$.

\section{System Model} \label{sec:system:model}
\subsection{General Description of the Communication Setup}

\par We consider a mmWave covert communication scenario, {where transmitter Alice wishes to communicate to receiver Bob, subject to the surveillance of warden Willie who attempts to detect the existence of this communication. It is assumed that Alice and Bob are equipped with Uniform Linear Arrays (ULA) of $N_a$ and $N_b$ antennas, respectively, so that directional transmission is possible between the two parties. In addition, both Alice and Bob deploy single RF chain to reduce hardware complexity as a commonly considered in existing works~\cite{wang2009beam,hur2013millimeter,liu2017Jsac,li2019explore}. As to warden Willie, he is always curious and greedy by nature and is thus assumed to deploy omni-directional antenna to monitor signal from all possible directions.}

\par {For the scenario described, we further assume a frame-slotted communication between Alice and Bob. As illustrated in Fig.~\ref{fig:sysmodel}, each frame has $n$ symbols in total (e.g., on the order of channel coherence time) and is further divided into a beam alignment (BA) phase that consists of $n_{a}$ symbols and a data transmission (DT) phase of $n-n_{a}$ symbols. In the BA phase, Alice and Bob jointly train transmit/receive beam pairs from pre-designed codebooks so as to determine the best beam pair that is then used for the subsequent DT phase. Both BA and DT phases should be carefully designed so that the directional link between Alice and Bob is sufficiently good and the probability of detection of communication at Willie is kept at the covertness level required.}

\par {In what follows, we first elaborate the signalling model for communication between Alice and Bob and then define the binary hypothesis detection problem at Willie.}

\subsection{Signalling Model for Beam Alignment and Data Transmission Between Alice and Bob}\label{subsec:signal:model:Alice:Bob}

\par {We assume that Alice and Bob adopt an exhaustive-search (ES) strategy for beam training~\cite{liu2017Jsac}. Specifically, let ${\mathcal C}_{a} =\{{\bar {\bf w}_{l_a}}\in {\mathbb C}^{N_a \times 1}, l_a \in [1:L_a]\}$ be a set of $L_a$ unit-norm beams that jointly cover the Angle of Departure (AoD) interval at Alice, while ${\mathcal C}_{b} =\{{\bar {\bf f}_{l_b}}\in {\mathbb C}^{N_b \times 1}, l_b \in [1:L_b]\}$ be a set of $L_b$ unit-norm beams that jointly cover the Angle of Arrival (AoA) interval at Bob. The entire training codebook is then formed by considering all possible $L = L_a L_b$ Alice/Bob beam pairs, i.e., ${\mathcal C} = \{({\bf w}, {\bf f}): {\bf w} \in {\mathcal C}_{a}, {\bf f} \in {\mathcal C}_{b}\}$.}

\par {For each $({{\bf w}_{l}, {\bf f}_{l}}) \in {\mathcal C}$, Alice sends a pilot sequence via beam ${\bf w}_{l}$, while Bob performs an output measurement via beam ${\bf f}_{l}$. The output signal at Bob is given by
\begin{align}
{\mathbf y}_{l}^{\text{p}}= \sqrt{P_a}{\mathbf f}_l^{\dag} {\mathbf H}_{ab} {\mathbf w}_l {\bf x}^{\text{p}} + {\mathbf z}_b^{\text{p}},~~l \in [1:L], \label{equ:y:output}
\end{align}
where $P_a$ is the transmit power for beam training at Alice, ${\mathbf x}^{\text{p}} \in {\mathbb C}^{n_p \times 1}$ denotes the pilot sequence with $\|\mathbf{x}^p\|_2^2 = n_p$, ${\mathbf H}_{ab} \in {\mathbb C}^{N_b \times N_a}$ denotes the channel between Alice and Bob, ${\mathbf z}_b^{\text{p}} \in {\mathbb C}^{n_p \times 1}$ is the equivalent channel noise vector after received beamforming, whose elements are i.i.d. as ${\mathbf z}_{b,i}^{\text{p}}\sim{\mathcal{CN}}(0,\sigma_b^2)$. Assuming that the beam pairs in ES are allocated with equal training budget, we thus have $n_p = n_a/L$.}

\par {In particular, we consider single-path light-of-sight (LOS) channel between Alice and Bob, and thus channel ${\mathbf H}_{ab}$ can be specialized to}
\begin{align}
{\mathbf H}_{ab} = \gamma_{ab} {\mathbf u}(\phi){\mathbf v}^{\dag}(\psi),
\end{align}
{where $\gamma_{ab}$ is the channel coefficient}, while ${\mathbf u}(\phi)\in {\mathbb C}^{N_b \times 1}$ and ${\mathbf v}(\psi) \in {\mathbb C}^{N_a \times 1}$ are the steering vectors corresponding to AoA $\phi$ and AoD $\psi$ that are defined as
\begin{align}
&{\mathbf u}(\phi)= [1, e^{j2\pi\frac{d}{\lambda}\sin(\phi)},\cdots,e^{j2\pi\frac{d}{\lambda}(N_R-1)\sin(\phi)}]^T,\\
&{\mathbf v}(\psi)= [1, e^{j2\pi\frac{d}{\lambda}\sin(\psi)},\cdots,e^{j2\pi\frac{d}{\lambda}(N_T-1)\sin(\psi)}]^T,
\end{align}
respectively, with $\lambda$ being the wave-length and $d$ being the antenna spacing. Under this model {and when $({\mathbf w}_l, {\mathbf f}_l)$ beam pair is trained, the effective channel in~\eqref{equ:y:output} is specialized to}
\begin{align}
h_l  =  \gamma_{ab} {\mathbf f}^{\dag}_l {\mathbf u}(\phi){\mathbf v}^{\dag}(\psi) {\mathbf w}_l, \label{equ:single:path}
\end{align}
with {beamforming gain}
\begin{equation}
\begin{split}
G_l &\triangleq |{\mathbf f}^{\dag}_l {\mathbf u}(\phi){\mathbf v}^{\dag}(\psi) {\mathbf w}_l|^2\\
&=F_l(\phi) W_l(\psi), \label{equ:gain:1}
\end{split}
\end{equation}
{where $W_l(\psi) \triangleq |{\mathbf v}^{\dag}(\psi) {\mathbf w}_l|^2$ is the transmit beamforming gain along AoD $\psi$ at Alice side, while $F_l(\phi) \triangleq |{\mathbf f}^{\dag}_l {\mathbf u}(\phi)|^2$ is the receive beamforming gain along AoA $\phi$ at Bob side.}

\begin{figure}[t]
	\centering
	\includegraphics[width=0.5\textwidth]{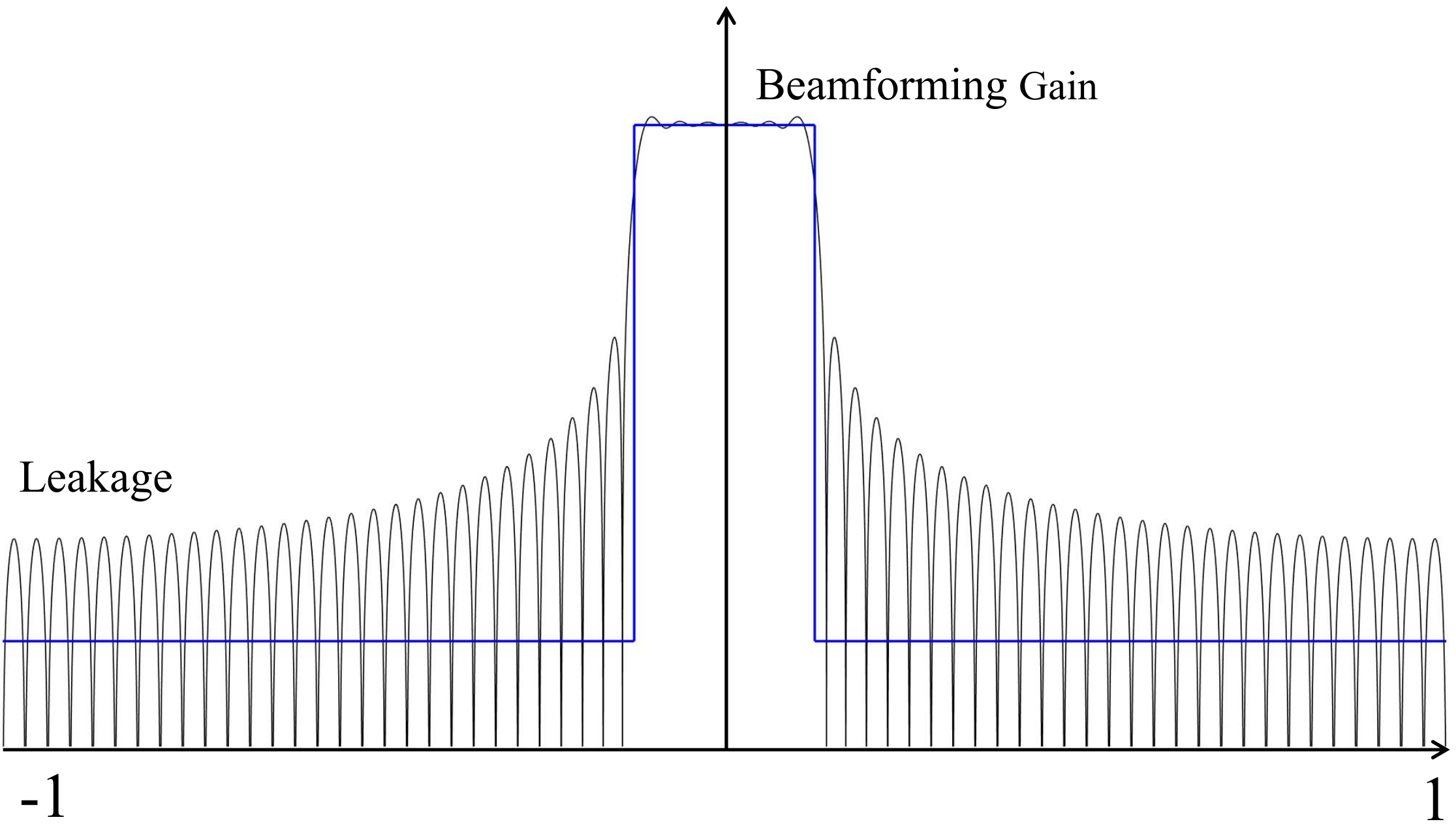}
	\caption{An illustration of the generalized flat-top beam pattern adopted.}
	\label{fig:beampattern}
	\vspace{-1em}
\end{figure}

\par {We also consider that each of the beams trained has uniform gain within its intended coverage interval (i.e., its mainlobe) and constant small leakage outside the mainlobe (as illustrated in Fig.~\ref{fig:beampattern}) as in \cite{7037401}. This slightly generalizes the commonly used flat-top beam model and is useful to capture the side-lobe leakage of non-ideal beams in practice. Moreover, assuming that all Alice (Bob) beams have equal-size non-overlapping mainlobes that jointly cover the AoD range $\Psi$ (resp. AoA $\Phi$), so Alice (Bob) beamforming gain can then be represented as
\begin{align}
\begin{split}
W_l(\psi)  = \left\{ {\begin{array}{*{20}c}
	W_a, ~~\text{if~}\psi \in \Psi_{{\bf w}_l} \\
	w_a,~~\text{otherwise}  \\
	\end{array}} \right.\\
\end{split}
\end{align}
and
\begin{align}
\begin{split}
F_l(\phi)  = \left\{ {\begin{array}{*{20}c}
	F_b, ~~\text{if~}\phi \in \Phi_{{\bf f}_l} \\
	f_b,~~\text{otherwise}  \\
	\end{array}} \right.,\label{equ:gain}
\end{split}
\end{align}
respectively, where $\Psi_{{\bf w}_l}$ and $\Phi_{{\bf f}_l}$ denote the mainlobe interval (in the $\textit{sin}$ domain) of beam ${\bf w}_l$ and ${\bf f}_l$, respectively, and $W_a \gg w_a$, $F_b \gg f_b$.}

\par {With the above assumptions, the output signal of~\eqref{equ:y:output} at Bob is then specialized to
\begin{align}
{\mathbf y}_{l}^{\text{p}}= \gamma_{ab}\sqrt{P_aG_l}{\bf x}^{\text{p}} + {\mathbf z}_b^{\text{p}},
\end{align}
where $G_l$ is drawn from the set $\{W_aF_b, w_aF_b, W_af_b, w_af_b\}$, which depends on how well the $l$th beam pair aligns with the underlying channel.}

\par {In addition, Alice and Bob are assumed to share the pilot sequences used for beam training beforehand. Given output measurement ${\mathbf y}_l^{\text{p}}$ and the known pilot sequence ${\bf x}^{\text{p}}$, Bob can then further form match-filtered outputs as
\begin{align}
\widetilde{y}_l^{\text{p}} &={({\mathbf x}^{\text{p}})^\dag{\mathbf y}_l^{\text{p}}}\\
&= n_p \gamma_{ab}\sqrt{P_aG_l} + ({\bf x}^{\text{p}})^{\dag}{\mathbf z}_b^{\text{p}},~~l \in [1:L].\label{equ:match:filtered:output}
\end{align}
The beam pair $({\mathbf w}_{\hat l_{\text{ES}}}, {\mathbf f}_{\hat l_{\text{ES}}})$ as leading to the strongest match-filtered output is then chosen the one used for subsequent data transmission, where ${\hat l_{\text{ES}}}$ is given by
\begin{align}
{\hat l}_{\text{ES}}  = \mathop {\arg \max }\limits_{l \in [1:L]} |\widetilde{y}_l^{\text{p}}|.\label{equ:exhaustive:search}
\end{align}}

\par {In this way, during the DT phase, the Alice-to-Bob channel input-output relationship is represented by
\begin{align}
\textbf{y}^{\text d}_b=\gamma_{ab}\sqrt{{P_d}G_d}\textbf{x}^{\text d}+\textbf{z}_{b}^{\text d}, \label{equ:yb}
\end{align}
where $\textbf{x}^{\text{d}} \in {\mathbb C}^{(n-n_a)\times 1}$ is the input data vector with i.i.d. elements $\sim {\mathcal{CN}}(0,1)$, $P_d$ is the transmit power for data communication, $G_d = G_{\hat l_{\text{ES}}}$ with $\hat l_{\text{ES}} $ as in~\eqref{equ:exhaustive:search}, $\textbf{z}_{b}^{\text d} \in {\mathbb C}^{(n-n_a)\times 1}$ noise vector with i.i.d. elements $\sim{\mathcal{CN}}(0,\sigma_b^2)$ and $\textbf{y}^{\text d}_b$ is the output signal at Bob.}

\par {It is clear that the beamforming gain $G_d$ in~\eqref{equ:yb} takes only one of $\{W_aF_b, W_af_b, w_aF_b, w_af_b\}$, which depends on the beam alignment performance via ES beam training. Motivated by this, we introduce a notion of average effective throughput
\begin{align}
\overline{T} = (1 - \frac{n_a}{n}){\mathbb E}\left\{\log \left(1+\frac{P_d|\gamma_{ab}|^2}{\sigma_b^2}G_d \right)\right\},\label{effective:SNR}
\end{align}
to measure the average performance of the Alice-Bob data link, which explicitly takes into account the impact of beam alignment overhead and accuracy on the subsequent data communication. Unless otherwise specified, the $\log$ function takes base 2.}

\subsection{Binary Detection Problem at Willie}
\par {In order to determine the presence of Alice-to-Bob covert communication, Willie needs to distinguish the following two hypotheses
\begin{align}
\left\{ {\begin{array}{*{20}l}
   {{\mathcal H}_0 :~{\bf{y}}_w  = {\bf{z}}_w }  \\
   {{\mathcal H}_1 :~{\bf{y}}_w  = {\bf{s}} + {\bf{z}}_w }  \\
\end{array}} \right.
 \label{equ:yw:signal:wiliie}
\end{align}
where all ${\bf{y}}_w, {\bf{s}}, {\bf{z}}_w \in {\mathbb C}^{n\times 1}$. Note that $\mathcal{H}_{0}$ denotes the null hypothesis where Alice has not communicated with Bob and thus only channel noise vector ${\bf{z}}_w$ (with i.i.d. elements $\sim {\mathcal{CN}}(0,\sigma^2_w)$) is observed, while $\mathcal{H}_{1}$ denotes the alternative hypothesis where Alice has communicated with Bob and thus some information leakage superimposed on channel noise is observed.}

\par {Under $\mathcal{H}_{1}$, to be more specific, considering that the Alice-to-Willie link is also in LOS, the resultant channel is then represented as
\begin{align}
{\mathbf h}_{aw} = \gamma_{aw}{\mathbf v}^{\dag}(\psi_{aw}),
\end{align}
where $\psi_{aw}$ and $\gamma_{aw}$ are the associated AoD and channel coefficient, respectively. The signal vector i.e., $\mathbf{s}$ at Willie is thus formed by two parts: The first $n_a$ symbols (i.e., ${\bf{s}}_{[1:n_a]}$) are the signals at Willie when Alice and Bob perform beam training
\begin{align}
{\bf s}_{[(l-1)*n_p+1:l*n_p]} &= \sqrt{P_a}{\mathbf h}_{aw}{\bf w}_{l}{\mathbf x}^{\text{p}} \nonumber\\
 &= \left\{ {\begin{array}{*{20}l}
	\gamma_{aw}\sqrt{P_aW_a}{\bf{x}}^{\text{p}},~{\textrm{if}}~\psi_{aw} \in \Psi_{{\bf w}_l} \\
	\gamma_{aw}\sqrt{P_aw_a}{\bf{x}}^{\text{p}},~\textrm{otherwise},\\
	\end{array}} \right. \label{equ:S1}
\end{align}
while the rest are the signals at Willie when Alice and Bob perform data transmission:
\begin{align}
{\bf s}_{[n_a+1:n]} &= \sqrt{P_d}{\mathbf h}_{aw}{\mathbf w}_{\hat l_{\text{ES}}}{\mathbf x}^{\text{d}}.\label{equ:S2}
\end{align}}

\par {Given ${\bf s}$, Willie makes a binary decision ($\mathcal{D}_{1}$ or $\mathcal{D}_{0}$) that infers whether Alice's transmission is present or not. Consider equal a priori probability of $ {\mathcal H}_0 $ and $ {\mathcal H}_1 $. To measure the detection performance of Willie, we adopt the total detection error probability $\xi$, which is defined as
\begin{align}
\xi=\alpha+\beta,
\end{align}
where $\alpha\triangleq \text{Pr}(\mathcal{D}_{1}|\mathcal{H}_{0}
)$ denotes the false alarm rate, $\beta\triangleq \text{Pr}(\mathcal{D}_{0}|\mathcal{H}_{1})$ denotes the missed detection rate. Let $\xi^*$ be the minimum error probability Willie can achieve by using an optimal detector.}

\par { Our ultimate goal is thus to develop appropriate Alice-to-Bob beam training and data transmission design so as to maximize $\overline{T}$ for Alice-Bob link, while enforcing that $\xi^\ast\geq1-\epsilon$ at Willie for a covertness level $\epsilon >0$ required. Towards this end, in what follows, we shall first characterize Alice-Bob $\overline{T}$ and Willie's detection performance as a function of key system parameters (including training duration $n_a$, transmit power $P_a$ and $P_d$ for BA and DT), and then propose a joint optimized design of BA and DT for the covert communication studied.}

\section{Joint Optimization of Beam Alignment and Data Transmission for Covert MmWave Communication}\label{sec:joint:optimization}

\subsection{{Characterization of $\overline{T}$ for Alice-Bob Link }} \label{sec:alice-to-bob:link}

\par {For the Alice-Bob link, recall from~\eqref{effective:SNR} that the average effective throughput $\overline{T}$ crucially depends on the statistical property of beamforming gain $G_d$ after beam training. In particular, $G_d= W_aF_b$ when there is perfect beam alignment, while $G_d$ takes a much smaller gain from the set $\{W_af_b, w_aF_b, w_af_b\}$ if there is one-sided or two-sided misalignment.}

\par {To quantify $\overline{T}$, we introduce the following probability of successful alignment through ES beam training
\begin{align}
p_{\text{align}} \triangleq \Pr\{ \hat l_{\textrm{ES}} = l_\textrm{opt}\},
\end{align}
where $l_\textrm{opt} = \mathop {\arg \max }\nolimits_{l \in [1:L]} |{\mathbf f}^{\dag}_l {\mathbf u}(\phi){\mathbf v}^{\dag}(\psi) {\mathbf w}_l|^2$ is the index of the optimal beam pair that leads to the largest beamforming gain. Considering the facts that beamforming gain is much larger under perfect alignment and that we ought to achieve high $p_{\text{align}}$, we can approximate $\overline{T}$ as
\begin{align}
\overline{T} \approx (1 - \frac{n_a}{n})\log \left(1+\frac{P_d|\gamma_{ab}|^2}{\sigma_b^2}W_aF_b\right)p_{\text{align}}\label{equ:T:eff}
\end{align}
by dropping the marginal throughput contribution in the case of misalignment for the sake of tractability.}

\par {We now further analyze $p_{\text{align}}$. Without loss of optimality and for notational convenience, $l_{\textrm{opt}} =1$ is assumed. Based on the ES beam training~\eqref{equ:exhaustive:search}, $p_{\text{align}}$ can be represented as
\begin{align}
p_{\text{align}} &= \Pr\{ \hat l_{\textrm{ES}} = l_\textrm{opt}\}  \nonumber \\
                 &= \Pr\{|\tilde y_1^{\textrm{p}}| > \max\{|\tilde y_2^{\textrm{p}}|, |\tilde y_3^{\textrm{p}}|,\cdots, |\tilde y_N^{\textrm{p}}|\}\} \\
                 &=  1 - \Pr\{|\tilde y_1^{\textrm{p}}| \le \max\{|\tilde y_2^{\textrm{p}}|, |\tilde y_3^{\textrm{p}}|,\cdots, |\tilde y_N^{\textrm{p}}|\}\} \\
                 & = 1 - \Pr\{T_1\le \max\{T_2, T_3,\cdots, T_L\}\}, \label{equ:p:align}
\end{align}
where we have defined normalized statistics
\begin{align}
T_l = \frac{|\tilde y_l^{\textrm{p}}|^2}{\frac{\sigma^2}{2}n_p},~~l\in [1:L]. \label{equ:normalised:T}
\end{align}}
\par {Let $\chi _k^2 \left( \lambda  \right)$ denote a noncentral chi-squared distribution with degrees of freedom (DoFs) $k$ and noncentral parameter $\lambda$. To derive useful properties of $p_{\text{align}}$, we introduce the following lemma.}

{\begin{lemma} \label{lemma:statistics}
The normalized statistics defined all follow noncentral chi-squared distribution with DoFs $k=2$ and with noncentral parameter drawn from the set $\{\lambda_A, \lambda_B, \lambda_C, \lambda_D\}$ defined as:
\begin{align}
\lambda_A &= \frac{2|\gamma_{ab}|^2n_pP_aW_aF_b}{\sigma_b^2},~~\lambda_B = \frac{2|\gamma_{ab}|^2n_pP_aw_aF_b}{\sigma_b^2},\\
\lambda_C &= \frac{2|\gamma_{ab}|^2n_pP_aW_af_b}{\sigma_b^2},~~\lambda_D = \frac{2|\gamma_{ab}|^2n_pP_aw_af_b}{\sigma_b^2}.
\end{align}
Specifically, $T_1 \sim \chi _2^2 \left(\lambda_A\right)$, while among $\{T_2, T_3,\cdots, T_L\}$, $(L_a-1)$ variables follow $\chi _2^2 \left(\lambda_B\right)$, $(L_b-1)$ variables follow $\chi _2^2 \left(\lambda_C\right)$ and $(L_a-1)(L_b-1)$ variables follow $\chi _2^2 \left(\lambda_D\right)$.
\end{lemma}
\begin{IEEEproof}
Following~\eqref{equ:match:filtered:output}, $T_l$ of~\eqref{equ:normalised:T} is represented by
\begin{align}
T_l &=  \frac{|\tilde y_l^{\textrm{p}}|^2}{\frac{\sigma^2}{2}n_p} = \frac{|n_p\gamma_{ab}\sqrt{P_aG_l} + ({\bf x}^{\text{p}})^{\dag}{\mathbf z}_b^{\text{p}}|^2}{\frac{\sigma^2}{2}n_p} \\
&= \left|\gamma_{ab}\sqrt{\frac{n_pP_aG_l}{\frac{\sigma^2}{2}}} + \frac{({\bf x}^{\text{p}})^{\dag}{\mathbf z}_b^{\text{p}}}{\sqrt{\frac{\sigma^2}{2}n_p}}\right|^2.
\end{align}
which implies that
\begin{align}
T_l \sim \chi _2^2 \left(\frac{2|\gamma_{ab}|^2n_pP_aG_l}{\sigma_b^2}\right),~~l\in[1:L],
\end{align}
by the definition of noncentral chi-squared distribution, where \textit{i}) when Alice and Bob's beams are perfectly aligned, $G_1 = W_aF_b$; \textit{ii}) when one-sided misalignment occurs at Alice or Bob side, $G_l = w_aF_b$ or $G_l = W_af_b$, respectively; \textit{iii}) when misalignment occurs at both Alice and Bob, $G_l = w_af_b$. Moreover, these variables are independent, since they are constructed from training measurements at different time.
\end{IEEEproof}}
\par {For ease of exposition and without loss of generality, we further assume that variables $\{T_2,\cdots, T_{L_a+1}\} \sim \chi _2^2 \left(\lambda_B\right)$, $\{T_{L_a+2},\cdots, T_{L_a+L_b+1}\} \sim \chi _2^2 \left(\lambda_C\right)$ and $\{T_{L_a+L_b+2},\cdots, T_{L}\} \sim \chi _2^2 \left(\lambda_D\right)$. We now establish a lower bound on $p_{\text{align}}$ as stated in the following proposition.}

\begin{proposition}\label{prop:align}
{The probability of successful alignment $p_{\text{align}}$ is lowered bound by ${p}_{\text{LB}}$:
\begin{align}
{p}_{\text{LB}}(P_a, n_p) = 1 - {p}_{\text{miss},1} -  {p}_{\text{miss},2} - {p}_{\text{miss},3},
\end{align}
where
\begin{align}
{p}_{\text{miss},1} &= 1-\int_{\rm{0}}^\infty  {\left( {F\left( {t\left| {2,\lambda _B } \right.} \right)} \right)^{L_a  - 1} f\left( {t\left| {2,\lambda _A } \right.} \right)dt}, \\
{p}_{\text{miss},2} &= 1-\int_{\rm{0}}^\infty  {\left( {F\left( {t\left| {2,\lambda _C } \right.} \right)} \right)^{L_b  - 1} f\left( {t\left| {2,\lambda _A } \right.} \right)dt},  \\
{p}_{\text{miss},3} &= 1-\int_{\rm{0}}^\infty  {\left( {F\left( {t\left| {2,\lambda _D } \right.} \right)} \right)^{(L_a-1)(L_b-1)} f\left( {t\left| {2,\lambda _A } \right.} \right)dt},
\end{align}
with $f\left( {t\left| {k,\lambda} \right.} \right)$ and $F\left( {t\left| {k,\lambda} \right.} \right)$ being the probability density function (pdf) and the cumulative distribution function (cdf) of $\chi _k^2 \left( \lambda  \right)$, respectively.}
\end{proposition}
\begin{IEEEproof}
{The proof uses the definition of $p_{\text{align}}$ as in~\eqref{equ:p:align} and the statistical properties of the random variables established in Lemma~\ref{lemma:statistics}. The details are deferred to Appendix~\ref{appendix:prop:palign}.}
\end{IEEEproof}

\par Based on this proposition and from~\eqref{equ:T:eff}, the approximated $\overline{T}$ is lowered bound by $\overline{T}_{\textrm{LB}}$ given by:
\begin{equation}
\begin{split}
\overline{T}_{\textrm{LB}} &= (1 - \frac{n_a}{n})\log \left(1+\frac{P_d|\gamma_{ab}|^2}{\sigma_b^2}W_aF_b\right){p}_{\text{LB}}(P_a, n_p) \\
&= (1 - \frac{n_pL}{n})\log \left(1+\frac{P_d|\gamma_{ab}|^2}{\sigma_b^2}W_aF_b\right){p}_{\text{LB}}(P_a, n_p) .\label{equ:throughput:LB}
\end{split}
\end{equation}

{
\subsection{{Detection Performance at Willie}}
\par As for Willie, let ${\mathbb P}_0$ and ${\mathbb P}_1$ be the probability distribution of its observations under $\mathcal{H}_{0}$ and $\mathcal{H}_{1}$ as in~\eqref{equ:yw:signal:wiliie}, respectively. In particular, under $\mathcal{H}_{0}$, the distribution ${\mathbb P}_0$ is given by:
\begin{align}
{\mathbb P}_0=\dfrac{1}{(\pi\sigma_w^2)^{n}}\text{exp}\left(-\dfrac{\sum_{i=1}^n\mid y_w^i\mid^2}{\sigma_w^2}\right),\label{equ:p0}
\end{align}
since only noises are observed at Willie. Under $\mathcal{H}_{1}$, the distribution ${\mathbb P}_1$ can be represented by:
\begin{align}
{\mathbb P}_1 = {\mathbb P}_1^{{\textrm{BA}}} \times {\mathbb P}_1^{{\textrm{DT}}},
\end{align}
where ${\mathbb P}_1^{{\textrm{BA}}}$ corresponds to the joint distribution of received signals at Willie when Alice and Bob are in the BA phase, while ${\mathbb P}_1^{{\textrm{DT}}}$ corresponds to the joint distribution of received signals at Willie when Alice and Bob are in the DT phase. Considering that Willie has no knowledge of the pilot sequence used for beam training between Alice and Bob, based on~\eqref{equ:yw:signal:wiliie} and~\eqref{equ:S1}, ${\mathbb P}_1^{{\textrm{BA}}}$ is approximatively characterized by
\begin{small}
\begin{align}
&{\mathbb P}_1^{{\textrm{BA}}}=\dfrac{1}{(\pi(\sigma_w^2+|\gamma_{aw}|^2P_aW_a))^{{L_bn_p}}}\text{exp}\left(-\dfrac{\sum_{i=1}^{n_p}\mid y_w^i\mid^2}{\sigma_w^2+|\gamma_{aw}|^2P_aW_a}\right)\nonumber\\
&\times\dfrac{1}{(\pi(\sigma_w^2+|\gamma_{aw}|^2P_aw_a))^{{L_b(L_a-1)n_p}}}\text{exp}\left(-\dfrac{\sum_{i=n_p+1}^{n_a}\mid y_w^i\mid^2}{\sigma_w^2+|\gamma_{aw}|^2P_aw_a}\right) \label{equ:pBA}.
\end{align}
\ignorespacesafterend
\end{small}
To characterize ${\mathbb P}_1^{{\textrm{DT}}}$, we note from~\eqref{equ:S2} that Willie's received signals would depend on whether it is in the main-lobe or in the side-lobe of Alice's chosen beam for data transmission. To account for this factor, let $\rho$ be the probability that Willie is in the main-lobe of Alice's data beam. Then ${\mathbb P}_1^{{\textrm{DT}}}$ can be characterized by a mixture of ${\mathbb P}_{1,1}^{{\textrm{DT}}}$ and ${\mathbb P}_{1,2}^{{\textrm{DT}}}$ as
\begin{align}
{\mathbb P}_1^{{\textrm{DT}}} = \rho {\mathbb P}_{1,1}^{{\textrm{DT}}} +  (1-\rho){\mathbb P}_{1,2}^{{\textrm{DT}}}, \label{equ:PDT:mixture}
\end{align}
where we have
\begin{small}
\begin{align}
&{\mathbb P}_{1,1}^{{\textrm{DT}}} = \dfrac{1}{(\pi(\sigma_w^2+|\gamma_{aw}|^2P_dW_a))^{{n-n_a}}}\text{exp}\left(-\dfrac{\sum_{i=n_a+1}^{n}\mid y_w^i\mid^2}{\sigma_w^2+|\gamma_{aw}|^2P_dW_a}\right), \nonumber \\
&{\mathbb P}_{1,2}^{{\textrm{DT}}} = \dfrac{1}{(\pi(\sigma_w^2+|\gamma_{aw}|^2P_dw_a))^{{n-n_a}}}\text{exp}\left(-\dfrac{\sum_{i=n_a+1}^{n}\mid y_w^i\mid^2}{\sigma_w^2+|\gamma_{aw}|^2P_dw_a}\right), \nonumber
\end{align}
\ignorespacesafterend
\end{small}
as the joint probability distribution of Willie's received signals when it is in the main-lobe and in
the side-lobe of Alice¡¯s data beam, respectively.

\par {Based on the ${\mathbb P}_0$ and ${\mathbb P}_1$ computed, Willie performs a binary hypothesis testing. It is known that the error probability that Willie can achieved is lower bounded by:
\begin{align}
\xi^\star=1-\nu_T({\mathbb P}_0,{\mathbb P}_1),
\end{align}
where $\nu_T({\mathbb P}_0,{\mathbb P}_1)$ is the total variation distance between ${\mathbb P}_0$ and ${\mathbb P}_1$. However, the closed-form expression of this total variation is hard to obtain for the given ${\mathbb P}_0$ and ${\mathbb P}_1$ in our case. Alternatively, as in many existing works~\cite{BashGT13,WangWZ16,yan2018delay}, we consider the following upper bound on the total variation by using Pinsker's inequlity\cite{csiszarinformation}:
\begin{align}
\nu_T({\mathbb P}_0,{\mathbb P}_1)\leq\sqrt{\frac{1}{2}\mathcal D({\mathbb P}_0||{\mathbb P}_1)}
\end{align}
where $ \mathcal D({\mathbb P}_0||{\mathbb P}_1) $ is the relative entropy of ${\mathbb P}_0$ to ${\mathbb P}_1$ defined by:
\begin{align}
\mathcal D({\mathbb P}_0||{\mathbb P}_1) = \int p_0(x)\text{ln}\frac{p_0(x)}{p_1(x)} dx.
\end{align}
As a result, a sufficient condition to ensure the covertness constraint $\xi^\star\geq1-\epsilon$ at Willie is that
\begin{align}
\mathcal D({\mathbb P}_0||{\mathbb P}_1)\leq2\epsilon^2.\label{equ:covertness:cons}
\end{align}
\par We further note that ${\mathbb P}_1$ in our case contains a mixture of multivariate Gaussian component ${\mathbb P}_1^{{\textrm{DT}}}$ as in \eqref{equ:PDT:mixture}, which renders closed-form expression for $\mathcal D({\mathbb P}_0||{\mathbb P}_1)$ still difficult. To make the problem more tractable, we further approximate ${\mathbb P}_1^{{\textrm{DT}}}$ by a joint distribution $\overline{{\mathbb P}}_1^{{\textrm{DT}}}$ as given by
\begin{small}
	\begin{align}
    \overline{{\mathbb P}}_1^{{\textrm{DT}}} = \dfrac{1}{(\pi(\sigma_w^2+|\gamma_{aw}|^2P_d\overline{w_a}))^{{n-n_a}}}\text{exp}\left(-\dfrac{\sum_{i=n_a+1}^{n}\mid y_w^i\mid^2}{\sigma_w^2+|\gamma_{aw}|^2P_d\overline{w_a}}\right),\label{equ:p:a}
	\end{align}
\ignorespacesafterend
\end{small}
whose underlying variables are i.i.d. Gaussian distributed with zero mean and variance $(\sigma_w^2+|\gamma_{aw}|^2P_d\overline{w_a})$ with $\overline{w_a}=\rho W_a+(1-\rho)w_a$. Namely, we approximate each underlying mixture Gaussian variable in ${\mathbb P}_1^{{\textrm{DT}}}$ of~\eqref{equ:PDT:mixture} with a single Gaussian variable of the same mean and variance. Letting $\overline{{\mathbb P}}_1 = {\mathbb P}_1^{{\textrm{BA}}} \times \overline{{\mathbb P}}_1^{{\textrm{DT}}}$, we thus approximate $\mathcal D({\mathbb P}_0||{\mathbb P}_1)$ by $\mathcal D({\mathbb P}_0||\overline{{\mathbb P}}_1)$, which can be derived in closed-form as:
\begin{align}
\begin{split}
\mathcal D({\mathbb P}_0||\overline{{\mathbb P}}_1)&=L_bn_p\left[\ln \left(1+ \xi_1\right)-\frac{\xi_1}{1+\xi_1}\right]\\
&+L_b(L_a-1)n_p\left[\ln \left(1+ \xi_2\right)-\frac{\xi_2}{1+\xi_2}\right]\\
&+(n-n_a)\left[\ln\left(1+\xi_3\right)-\frac{\xi_3}{1+\xi_3}\right],\label{equ:d01}
\end{split}
\end{align}
with $\xi_1 ={|\gamma_{aw}|^2P_aW_a}/{\sigma_w^2}$, $\xi_2 = {|\gamma_{aw}|^2P_aw_a}/{\sigma_w^2}$ and $\xi_3 ={|\gamma_{aw}|^2P_d\overline{w_a}}/{\sigma_w^2}$.}
We note that this approximation is accurate in particular when the signal-to-noise-ratio (SNR) at Willie is relatively small (a typical case in covert communications).}
\setcounter{equation}{47}
\begin{figure*}[b]
	\hrulefill
	\begin{align}
	f(\bm{\theta})=\log\Big(1-\frac{n_pL}{n}\Big)+\log{p}_{\text{LB}}(P_a, n_p) +\log\log \Big(1+\frac{P_d|\gamma_{ab}|^2}{\sigma_b^2}W_aF_b\Big),
	\label{eq:ObjTransformP}
	\end{align}
\end{figure*}

\setcounter{equation}{45}
\subsection{{Problem Formulation}}
\par {Given the lower bound $\overline{T}_{\textrm{LB}}$ of~\eqref{equ:throughput:LB} on the throughput of Alice-Bob link and the covertness constraint of~\eqref{equ:covertness:cons} at Willie, we formulate the following optimization problem in our covert communication design:
\begin{subequations} \label{equ:optm:o}
\begin{align}
&\max_{P_a,P_d,n_p} (1 - \frac{n_pL}{n})\log \Big(1+\frac{P_d|\gamma_{ab}|^2}{\sigma_b^2}W_aF_b\Big){p}_{\text{LB}}(P_a, n_p) \\
&~~~~\text{s.t.}~~~\mathcal D({\mathbb P}_0||\overline{{\mathbb P}}_1)\leq2\epsilon^2,  \label{eq:problemC1}\\
&~~~~~~~~~~~ 1\le n_p \le \left \lfloor {\frac{n}{L}}\right \rfloor, n_p \in {\mathbb N},\label{eq:problemC2}
\end{align}
\end{subequations}
which aims to optimize the number of symbols $n_p$ allocated to each beam pair trained ($n_pL$ thus reflects the total training overhead), training power $P_a$ and data transmission power $P_d$ in order to maximize $\overline{T}_{\textrm{LB}}$, subject to the covertness constraint at Willie.}

\par {It is noted that while optimizing lower bound $\overline{T}_{\textrm{LB}}$ might not give exactly the same results as optimizing the true effective throughput, this problem still provide valuable insights into the tradeoff between beam training overhead and achievable rate for Alice-Bob link, and also the tradeoff between the rate performance of Alice-Bob link and the achievable covertness level against Willie. Specifically, spending more symbols on beam training would improve the beam alignment performance between Alice and Bob, but at the expense of reducing the time left for data transmission. In addition, having larger training power and data transmission power would improve the effective throughput of Alice-Bob link, but at the risk of violating the covertness constraint imposed on Willie.}

\par Solving problem \eqref{equ:optm:o} is quite challenging because the optimization variables are coupled in the nonconvex objective function and constraints. Moreover, the training overhead for each beam pair $n_p$ is a discrete variable, which further complicates the solution of problem \eqref{equ:optm:o}. Hence, we are faced with a mixed-integer nonlinear programming problem, which is usually considered as NP-hard. In principle, one can attempt to perform exhaustive search over variable space $(n_p,P_a,P_d)$ to find the optimal solution, but this would require traversing all possible $n_p$ values and proper discretization of $(P_a,P_d)$, which leads to extremely high computational complexity and an unaffordable computation overhead. In the next section, we shall propose a more efficient algorithm to solve this problem.

{
\section{Dual-Decomposition Successive Convex Approximation Algorithm}

In this section, we develop an efficient double-loop iterative algorithm named DSCA, which integrates dual-decomposition \cite{DualDec} with successive convex approximation (SCA) method \cite{SCA} to find the stationary solution of problem \eqref{equ:optm:o}. Specifically, we first recast problem \eqref{equ:optm:o} into a more tractable yet equivalent form by exploiting its structural properties. We then elaborate the design of the proposed algorithm and also prove its convergence to a local stationary point.

\subsection{Problem Reformulation}
Before proceeding to the derivation of the proposed algorithm, a suitable transformation for problem \eqref{equ:optm:o} is necessary, and we provide the following corollary:
\begin{corollary}\label{col:trans}
Problem \eqref{equ:optm:o} is equivalent to
\begin{align} \label{eq:transformedP}
&\max_{\bm{\theta}}~ f(\bm{\theta}) \\
&~~\text{s.t.}~~\eqref{eq:problemC1}-\eqref{eq:problemC2}\nonumber
\end{align}
where $\bm{\theta}\triangleq [P_a,P_d,n_p]^T$ denotes the composite optimization variable, and the objective function $f(\bm{\theta})$ is defined in \eqref{eq:ObjTransformP} as displayed at the bottom of the next page.
\end{corollary}

Since the $\log$ function is monotonically nondecreasing and also analytic in the real region, Corollary~\ref{col:trans} can be easily proved. It is noteworthy that the above equivalent transformation would facilitate the separation of optimization variables, thereby simplifying the subsequent development of the proposed algorithm.

{\setcounter{equation}{55}
\begin{figure*}[b]
	\hrulefill
	\begin{align}
	g(P_a)=\log{p}_{\text{LB}}(P_a) -\nu L_bn_p\left[\ln \left(1+ \xi_1\right)-\frac{\xi_1}{1+\xi_1}\right]-\nu L_b(L_a-1)n_p\left[\ln \left(1+ \xi_2\right)-\frac{\xi_2}{1+\xi_2}\right],
	\label{eq:ObjsubP_PA}
	\end{align}
\end{figure*}
\setcounter{equation}{62}
\begin{figure*}[b]
	\hrulefill
	\begin{align}
	g(n_p)=	&\log\Big(1-\frac{n_pL}{n}\Big)+\log{p}_{\mathrm{LB}}(n_p)-\Big\{L_b\Big[\ln \Big(1+ \xi_1\Big)-\frac{\xi_1}{1+\xi_1}\Big]\nonumber\nonumber\\
	&+ L_b(L_a-1)\Big[\ln \Big(1+ \xi_2\Big)-\frac{\xi_2}{1+\xi_2}\Big] -L_aL_b\Big[\ln \Big(1+ \xi_3\Big)-\frac{\xi_3}{1+\xi_3}\Big]\Big\}\nu n_p,
	\label{eq:ObjsubP_NP}
	\end{align}
\end{figure*}
\setcounter{equation}{64}

\begin{figure*}[hb]
	\hrulefill
	\begin{align}
	g_{c}(n_p)=	&-\Big\{L_b\Big[\ln \Big(1+ \xi_1\Big)-\frac{\xi_1}{1+\xi_1}\Big]+ L_b(L_a-1)\Big[\ln \Big(1+ \xi_2\Big)-\frac{\xi_2}{1+\xi_2}\Big]\nonumber\\
	& -L_aL_b\Big[\ln \Big(1+ \xi_3\Big)-\frac{\xi_3}{1+\xi_3}\Big]\Big\}\nu n_p+\log\Big(1-\frac{n_pL}{n}\Big),
	\label{eq:sub_NPobjc}
	\end{align}	
\end{figure*}

\setcounter{equation}{67}
\begin{figure*}[b]
	\hrulefill
	\begin{align}
	g(P_d)&=\log\log \Big(1+\frac{P_d|\gamma_{ab}|^2}{\sigma_b^2}W_aF_b\Big)-\nu(n-n_a)\left[\ln\left(1+\xi_3\right)-\frac{\xi_3}{1+\xi_3}\right],
	\label{eq:ObjsubP_PD}
	\end{align}
\end{figure*}}


\setcounter{equation}{48}
To make the problem tractable, we subsequently relax the discrete integer constraint \eqref{eq:problemC2} into a closed connected subset of the real axis, i.e., $1\le n_p \le \left \lfloor {\frac{n}{L}}\right \rfloor$. We remark that the limiting point generated by the proposed DSCA algorithm may not satisfy the integer constraint in \eqref{eq:problemC2}. To obtain an integer solution for the optimal training overhead required by each beam pair, we use the same method as in \cite{THCF} to round $n_p$ to its nearby integer as follows
\begin{equation}
	n_p(\delta)=\left\{
	\begin{aligned}
	\overset{} \lfloor{n_p^{\star}}\rfloor&,
	~~~~~\text{if}~n_p^{\star}-\lfloor{n_p^{\star}}\rfloor \leq \delta  \\
	\lceil{n_p^{\star}}\rceil&,~~~~~\text{otherwise,}\\
	\end{aligned}
	\right.
	\end{equation}
where $0\leq \delta \leq 1$ is chosen such that constraint \eqref{eq:problemC1} is met. Since $D({\mathbb P}_0||\overline{{\mathbb P}}_1)$ is monotonic in $n_p$, we can always find such $\delta$ using a bisection search over $\delta \in [0,1]$.
\setcounter{equation}{49}

However, problem \eqref{eq:transformedP} remains to be solved due to the nonlinear coupling of variables in the covertness constraint \eqref{eq:problemC1}. To overcome this difficulty, we decouple problem \eqref{eq:transformedP} into a master dual problem and a subproblem by leveraging dual decomposition method  \cite{DualDec}.
For the problem at hand, the first step is to introduce the nonnegative Lagrange multiplier $\nu$ associated with the covertness constraint \eqref{eq:problemC1} and to write the partial Lagrangian of problem \eqref{eq:transformedP} as
\begin{align}\label{eq:dualObj}
\mathfrak{L}(\bm{\theta},\nu)= f(\bm{\theta})-\nu\left(\mathcal D({\mathbb P}_0||\overline{{\mathbb P}}_1)-2\epsilon^2\right).
\end{align}
Then the dual function can be expressed as
\begin{align}
h(\nu)=\max_{\bm{\theta}}\mathfrak{L}(\bm{\theta},\nu)~~\text{s.t.}~~\eqref{eq:problemC1}.
\end{align}

Given the fixed dual variable $\nu$, the subproblem w.r.t. $\bm{\theta}$ can be written as
\begin{align}\label{SlaveP}
\max_{\bm{\theta}}~ f(\bm{\theta})-\nu\mathcal D({\mathbb P}_0||\overline{{\mathbb P}}_1) ~~\text{s.t.}~~\eqref{eq:problemC1}.
\end{align}
Based upon the optimal solution $\bm{\theta}^{\star}$ to problem \eqref{SlaveP}, we have the master dual problem in charge of updating the dual variable $\nu$ by solving the following problem:
\begin{align}\label{MasterP}
\min_{\nu\geq0} h(\nu)
\end{align}

In a nutshell, the master dual problem sets the price for the resources in each subproblem, which in turns decides the amount of used resources depending on the price \cite{Tolli_TWC2011}.
In the sequel, we show how the master dual problem \eqref{MasterP} and the subproblem \eqref{SlaveP} are efficiently solved. Hereinafter, we let superscript $t$ denote variables associated with the $t$-th iteration.


%
\subsection{Proposed DSCA Algorithm}
\emph{1) Subgradient Method for Solving Master Dual Problem}:
The Lagrangian duality theory in optimization states that the updating of the dual variable can be achieved via the minimization of $h(\nu)$, e.g., using the subgradient method \cite{CVX_Boyd}. Since the subgradient of $h$ at $\nu$ is simply the covertness constraint residual $\mathcal D({\mathbb P}_0||\overline{{\mathbb P}})-\epsilon^2$, \eqref{MasterP} can be efficiently solved with the following updates
\begin{equation}\label{eq:updateNu}
\nu^{t+1}=\left[\nu^t+\eta^t\left( \mathcal D^t({\mathbb P}_0||\overline{{\mathbb P}}_1)-2\epsilon^2
\right)
\right]_{+}
\end{equation}
where $\eta^t$ is the step size, $[\cdot]_{+}$ refers to the projection operation onto the nonnegative orthant, and $D^t({\mathbb P}_0||\overline{{\mathbb P}}_1)$ is the relative entropy of ${\mathbb P}_0$ to $\overline{{\mathbb P}}_1$ at the current point $\bm{\theta}^t$ generated by the last iteration.

Note that since the master dual problem \eqref{MasterP}  is always convex, the subgradient method is guaranteed to converge exactly to its globally optimal solution. In addition, the convergence properties of the subgradient method heavily rely on the choice of step size sequence $\{\eta^t\}$.  Constant step size is a typical choice of  $\{\eta^t\}$, but different step size should be properly chosen for different covertness level $2\epsilon^2$ for better performance. In this paper we adopt the diminishing step sizes as suggested in \cite{RTHP}.

\emph{2) IBCD Method for Solving Subproblem}: To solve subproblem \eqref{SlaveP}, we observe first that the constraints are separable with respect to the three blocks of variables, i.e., $P_a$, $P_d$, and $n_p$, which allows applying the inexact block coordinate descent (IBCD) method. Similar to the BCD method, the IBCD method sequentially updates each block of variables while fixing the other blocks to their previous values. Nevertheless, in the IBCD method, it is only required to find an inexact solution to some subproblems while keeping the objective function nondecreasing, rather than globally solving all the subproblem. For subproblem \eqref{SlaveP}, this amounts to the following steps:

\setcounter{equation}{54}


\textbf{Step 1(Updating $P_a$ While Fixing $P_d$ and $n_p$)}:  Let us now consider the subproblem w.r.t. $P_a$, which is given by
\begin{align}\label{eq:subP_PA}
\max_{P_a} ~ g(P_a)
\end{align}
where the objective function $g(P_a)$ is defined in \eqref{eq:ObjsubP_PA} as displayed at the bottom of the next page. Note that we cannot obtain the optimal $P^{\star}_a$ by directly maximizing $ g(P_a) $ due to the nonconcave objective function.  To address the challenge,  we first rewrite the objective function $ g(P_a) $ as follows:
\setcounter{equation}{56}
\begin{equation}
g(P_a)=g_{c}(P_a)+g_{n}(P_a),
\end{equation}
where $g_{c}(P_a)$ is the concave part of original objective function $g(P_a)$ given by
\begin{equation}
g_c(P_a)= \nu L_bn_p\frac{\xi_1}{1+\xi_1}+\nu L_b(L_a-1)n_p \frac{\xi_2}{1+\xi_2},\label{eq:subP_PAC}
\end{equation}
and  $g_{n}(P_a)$ is the nonconcave part of original objective function $g(P_a)$ given by
\begin{align}
g_n(P_a)=&\log{p}_{\text{LB}}(P_a) -\nu L_bn_p\ln \left(1+ \xi_1\right)\nonumber\\
&-\nu L_b(L_a-1)n_p\ln \left(1+ \xi_2\right).\label{eq:subP_PAN}
\end{align}
Then we preserve the partial concavity of the original objective function and linearize the nonconcave part, to construct the concave surrogate function $\hat{g}^{t}(P_a)$, resulting in the following
\begin{align}
\hat{g}^{t}(P_a)=g_{c}(P_a)+\Gamma_A^t(P_a-P^t_a)-\tau_{A}(P_a-P^t_a)^2,
\end{align}
where $\Gamma_A^t=\frac{\partial g_{n}^t(P^t_a)}{\partial P_a}$ is the partial derivative of $g_{n}^t(P_a)$ w.r.t. $ P_a$, and $\tau_A$ is a positive constant so that the surrogate function $\hat{g}^{t}(P_a)$ is strongly concave. Therefore, finding the optimal training power at the current iteration reduces to solving the following concave optimization problem:
\begin{align}\label{eq:updatePA}
{P}^{t+1}_a=\arg \max_{P_a} \hat{g}^{t}(P_a),
\end{align}
which can be efficiently solved by the the convex programming toolbox CVX \cite{CVX_Boyd}.

Here, we remark that the well-design surrogate function $\hat{g}^{t}(P_a)$ will help to speed up the convergence speed by preserving the structure of the original problem.
In addition, the presence of proximal regularization term $\tau_{A}(P_a-P^t_a)^2$ would further guarantee the algorithm convergence and can be properly chosen to achieve a good tradeoff between accuracy and computational complexity \cite{SCA}.

\textbf{Step 2(Updating $n_p$ While Fixing $P_d$ and $P_a$)}:  Let us now consider the subproblem w.r.t. $n_p$, which is given by
\begin{align}\label{eq:subP_NP}
\max_{1\le n_p \le \left \lfloor {\frac{n}{L}}\right \rfloor} ~ g(n_p)
\end{align}
where the objective function $g(n_p)$ is defined in \eqref{eq:ObjsubP_NP} as displayed at the bottom of this page.

Compared with problem \eqref{eq:ObjsubP_PA}, the above problem differs in that the training budget constraint on $n_p$ as in \eqref{eq:problemC2} is also considered. Following the same approach as used for updating $P_a$, we first decompose  objective function $ g(n_P) $ into the concave and nonconcave part and subsequently enable a SCA  of the original nonconvex problem as
\setcounter{equation}{63}
\begin{align}
\hat{g}^{t}(n_p)=g_{c}(n_p)+\Gamma_B^t(n_p-n^t_p)-\tau_{B}(n_p-n^t_p)^2,
\end{align}
where
$g_c(n_p)$ is the concave part of original objective function $g(n_p)$ is defined in \eqref{eq:sub_NPobjc} as given at the bottom of next page. Function $g_{n}(n_p)=\log{p}_{\text{LB}}(n_p)$ is the nonconcave part of original objective function $g(n_p)$;
and $\Gamma_B^t=\frac{\partial g_{n}^t(n^t_p)}{\partial n_p}$ is the partial derivative of $g_{n}^t(n_p)$ w.r.t. $ n_p$; $\tau_B$ is a positive constant so that the surrogate function $\hat{g}^{t}(n_p)$ is strongly concave. By applying the projection gradient method \cite{CVX_Boyd}, the optimal $n_p$ can be expressed as\setcounter{equation}{65}
\begin{align}\label{eq:updateNP}
n_p^{t+1}=\left[ \hat{n}^t_p\right]_{1}^{  \left \lfloor {\frac{n}{L}}\right \rfloor}
\end{align}
where $ \hat{n}^t_p=\arg \max_{n_p} \hat{g}^{t}(n_p)$ and $[\cdot]^{a}_{b}$ refers to the projection operation onto the box feasible region $[b,a]$.

\textbf{Step 3(Updating $P_d$ While Fixing $n_p$ and $P_a$)}: The variable $P_d$ is updated by solving the following unconstrained optimization problem:
\begin{align}\label{eq:subP_PD}
\max_{P_d} ~ g(P_d)
\end{align}
where the objective function $g(P_d)$ is defined in \eqref{eq:ObjsubP_PD} as displayed at the bottom of next page.
Likewise, we tailor a surrogate function of $g(P_d)$ with the specific
structure as follows:\setcounter{equation}{68}
\begin{align}
\hat{g}^{t}(P_d)=g_{c}(P_d)+\Gamma_C^t(P_d-P^t_d)-\tau_{C}(P_d-P^t_d)^2,
\end{align}
where $g_{c}(P_d)=\log\log \Big(1+\frac{P_d|\gamma_{ab}|^2}{\sigma_b^2}W_aF_b\Big)+\frac{\nu(n-n_a)\xi_3}{1+\xi_3}
$ is the concave term in $g_{c}(P_d)$, $g_{n}(P_d)=-\nu(n-n_a)\ln(1+\xi_3)$ is the rest nonconcave term, and $\Gamma_C^t=\frac{\partial g_{n}^t(P^t_d)}{\partial P_d}$ is the partial derivative of $g_{n}^t(P^t_d)$ w.r.t. $P_d$, and $\tau_C$ is a positive constant to be chosen.

Finding the optimal data transmission power is reduced to solving the following concave approximated problem:
\begin{align}
P^{t+1}_d=\arg\max_{P_d}~\hat{g}^{t}(P_d) \label{eq:updatePD}
\end{align}
which can be optimally determined by the standard convex optimization method.

\emph{3) Overall DSCA Algorithm}: According to the aforementioned results,  we summarize the proposed DSCA algorithm in Algorithm \ref{alg:DSCA}, where Steps 2-4 correspond to the three sub-iterations of the IBCD method. In subsequent, we establish the local convergence of the proposed DSCA algorithm to stationary solutions, using the following proposition.
\begin{proposition}\label{tho:Convergence}
The proposed algorithm produces  non-descending objective value sequence. Moreover, any limiting point $(P^{\star}_a,n^{\star}_p,P^{\star}_d)$ generated by the DSCA algorithm is a KKT point of problem \eqref{equ:optm:o}.
\end{proposition}
\begin{IEEEproof}
Refer to Appendix \ref{appendix:prop:cov} for the detailed proof.
\end{IEEEproof}
\par This proposition indicates that the proposed algorithm monotonically converges to a stationary point of problem \eqref{equ:optm:o}.  The monotonic convergence is attractive since it guarantees an improved objective value with arbitrary random initialization. In the next section, we will provide numerical results on the problem and draw insights on how much training is needed for the joint design of beam alignment and data transmission in covert mmWave communication.

\begin{algorithm}[t]
	\textbf{Initialization}:Define the tolerance of accuracy $\omega$ and the maximum number of iterations $T_{\mathrm{max}}$.  Let $t=0$ and choose a feasible initial point\;
	\Repeat{the increment on the value of the
		objective function in \eqref{eq:dualObj} is less than some threshold $\epsilon_2>0$ or reaching the maximum iteration number}{
		\textbf{Step 1}: Update $\nu^t$ according to \eqref{eq:updateNu}\;
		\textbf{Step 2}:  Update $P^t_a$ according to \eqref{eq:updatePA}\;
		\textbf{Step 3}:  Update $n_p^t$ according to \eqref{eq:updateNP}\;
		\textbf{Step 4}:  Update $P^t_d$ according to \eqref{eq:updatePD}\;
		\textbf{Step 5}:   Let $t=t+1$\;
	}
	\caption{Proposed DSCA Algorithm}
	\label{alg:DSCA}
\end{algorithm}
}

\section{Numerical Results}\label{sec:numerical:results}

\par {In the simulations below, Alice has $N_a = 32$ transmit antennas and Bob has $N_b = 8$ receive antenna. Unless stated otherwise, Alice and Bob are assumed to deploy $L_a = 32$ and $L_b =8$ generalized flat-top beams as defined in  Section~\ref{subsec:signal:model:Alice:Bob} to cover AoD and AoA $[0, 2\pi]$ range, respectively. For each beam, its main-lobe gain is assumed to be $0.5$ dB lower than that of ideal flat-top beam, and thus its side-lobe gain can be determined accordingly by the law of energy conservation. Therefore, we have that
{\small \begin{align}
&W_a=L_a 10^{-0.05},~~w_a=(2-(2/L_a*W_a))/(2-2/L_a);\\
&F_b=L_b10^{-0.05},~~f_b=(2-(2/L_b*F_b))/(2-2/L_b),
\end{align}}
for Alice and Bob's beams, respectively.}

\par {Each frame is assumed to have $n = 5120$ symbols in total for beam alignment and data transmission. This frame duration is on the order of the channel coherence time for low-mobility users in a mmWave system that operate at $73$~GHz with $100$-MHz bandwidth~\cite{liu2017Jsac}. Since the total number of beam pairs to be trained is $L = L_aL_b = 256$, the number of symbols that can be spent on each beam trained is up to $n/L= 20$. As for Alice-Bob and Alice-Willie links, we define $\kappa_b\triangleq|\gamma_{ab}|^2/\sigma_b^2$ and $\kappa_w\triangleq|\gamma_{aw}|^2/\sigma_w^2$ as the pre-beamforming SNR, respectively, which encapsulate the path-loss of the links. Moreover, consider that Willie is chosen uniformly at random in the beam space $[-1, 1]$ and therefore the probability of Willie being in the main-lobe of Alice's data beam is $\rho =1/L_a$ in the optimization problem~\eqref{equ:optm:o}.}
\begin{figure}[htbp]
	\centering
	\includegraphics[width=0.5\textwidth]{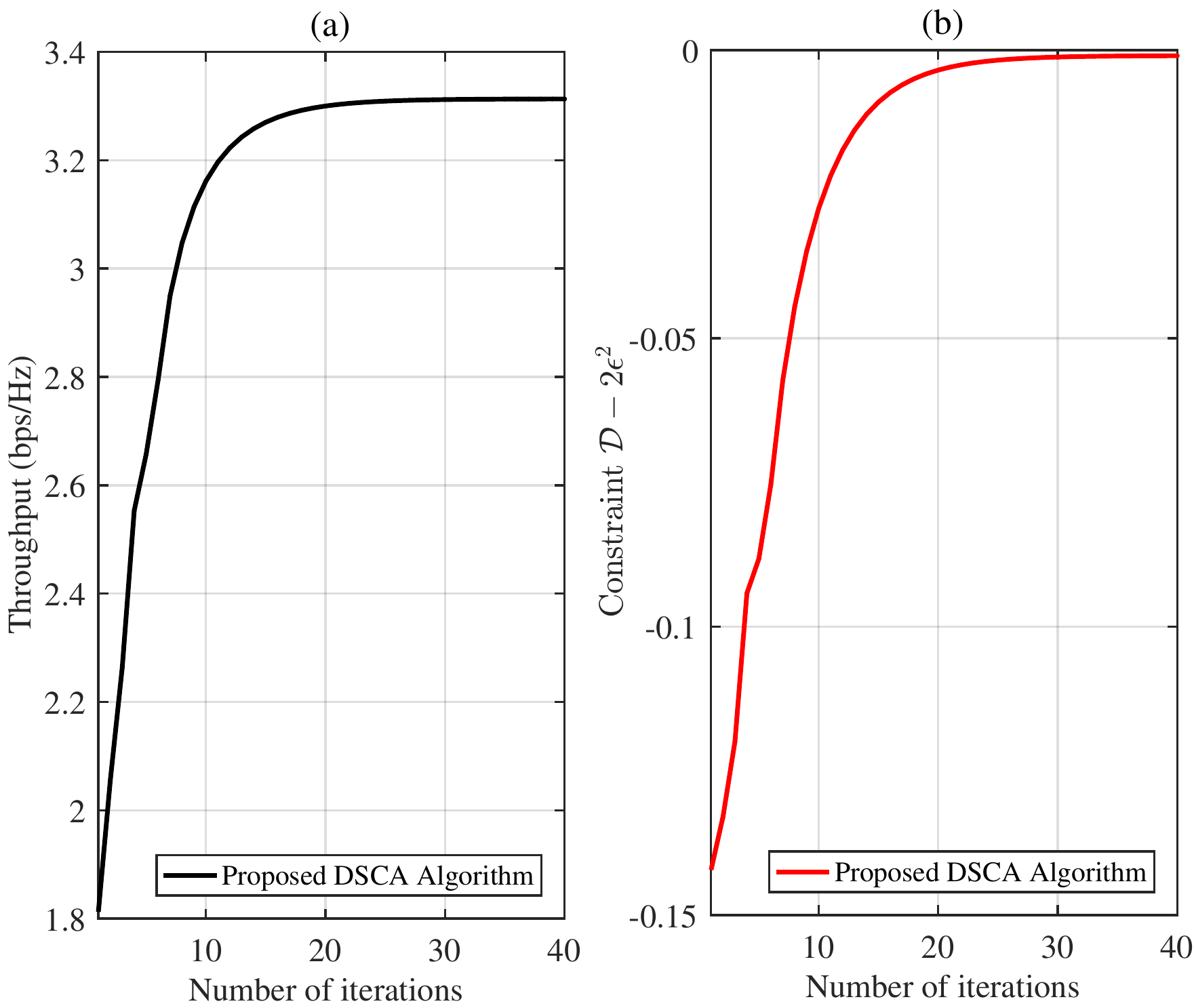}
	\caption{Convergence behavior of the DSCA algorithm.}
	\label{fig:converge}
\end{figure}

\par We first investigate the convergence behavior of the proposed DSCA algorithm. Consider that $\kappa_b = -5$ dB and $\kappa_w = -15$ dB and $\epsilon = 0.3$. Fig.~\ref{fig:converge} (a) and (b) plot an instance of the objective function and the constraint $\mathcal D({\mathbb P}_0||\overline{{\mathbb P}}_1) - 2\epsilon^2$ versus the number of iterations, respectively. It can be seen that the proposed DSCA algorithm quickly converges within a few tens of iteration and the optimal solution found by the DSCA algorithm satisfies the constraint of~\eqref{eq:problemC1} at strict equality. This demonstrates the efficiency of the proposed algorithm.

\par
\begin{figure}[htbp]
	\centering
	\includegraphics[width=0.45\textwidth]{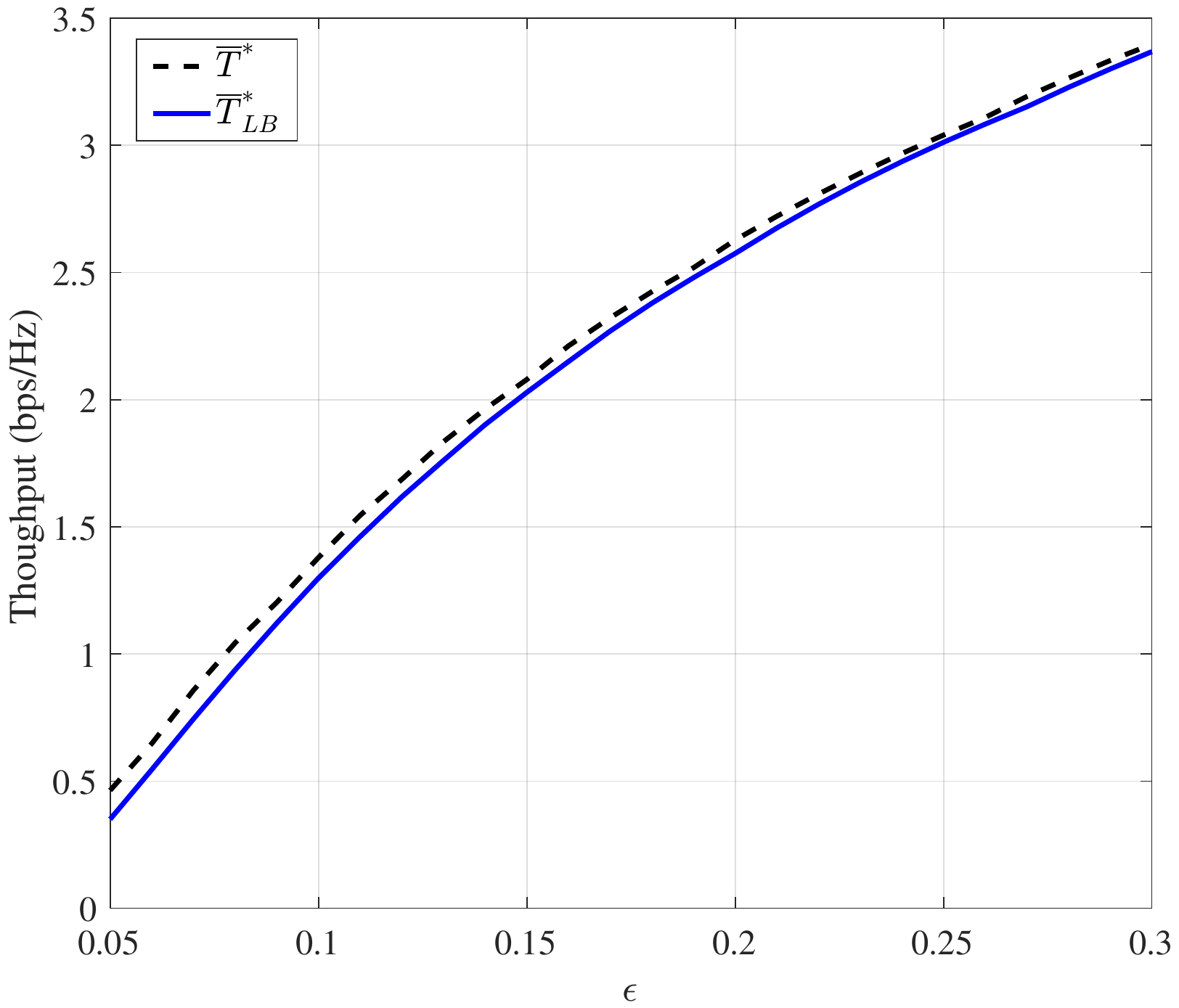}
	\caption{Throughput versus $ \epsilon $ ($\kappa_b=-5\text{dB}, \kappa_w=-15\text{dB}$).}
	\label{fig:TLB:1}
\end{figure}

\begin{figure}[htbp]
	\centering
	\includegraphics[width=0.45\textwidth]{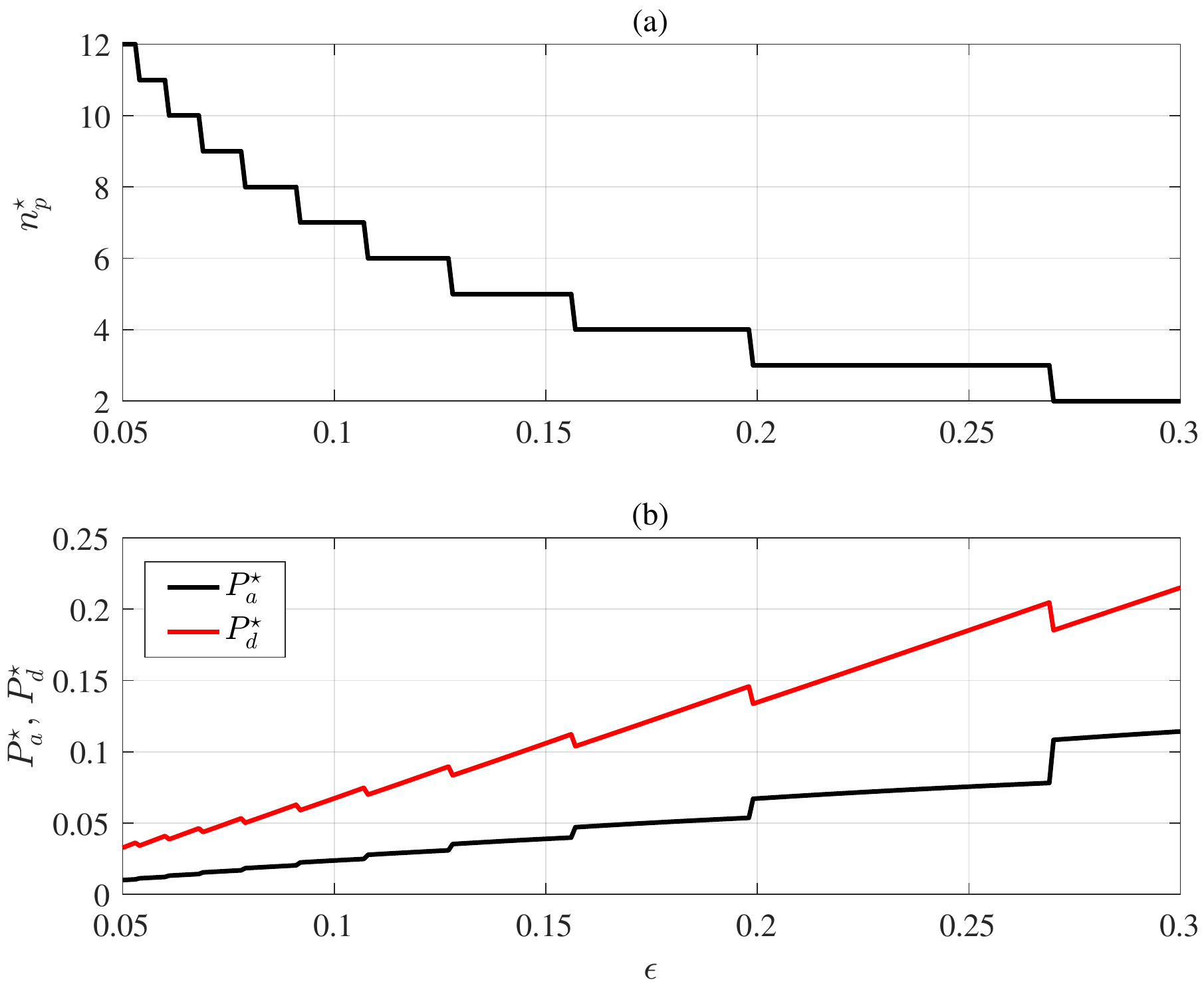}
	\caption{The optimal $(P_a^\star, P_d^\star, n_\textrm{p}^\star)$ versus $ \epsilon $ ($\kappa_b=-5\text{dB}, \kappa_w=-15\text{dB}$).}
	\label{fig:Np:pa:pd:1}
\end{figure}

 {Next, again fixing $\kappa_b = -5$ dB and $\kappa_w = -15$ dB, Fig.~\ref{fig:TLB:1} plots the optimized $\overline{T}_{\textrm{LB}}^\star$ by solving~\eqref{equ:optm:o} for different covertness levels $\epsilon \in [0.05:0.05:0.3]$ at Willie, while Fig.~\ref{fig:Np:pa:pd:1} shows the corresponding $(P_a^\star, P_d^\star, n_\textrm{p}^\star)$ that achieve $\overline{T}_{\textrm{LB}}^\star$ for each $\epsilon$ considered. In addition, with each $(P_a^\star, P_d^\star, n_\textrm{p}^\star)$ obtained, we also evaluate the approximate throughput $\overline{T}^\star$ in~\eqref{equ:T:eff} by computing $p_{\text{align}}$ via Monte-Carlo simulation and compare it against $\overline{T}_{\textrm{LB}}^\star$. It can be seen that $\overline{T}_{\textrm{LB}}^\star$ is generally close to $\overline{T}^\star$, which confirms the impact of the approximation in~\eqref{equ:throughput:LB} is small. In general, both $\overline{T}_{\textrm{LB}}^\star$ and $\overline{T}^\star$ get boosted when the covertness becomes less restricted (i.e., $\epsilon$ increases). This is reasonable since larger training power and data transmission power can be used at Alice to improve the beam alignment performance and also the SNR for data transmission as $\epsilon$ increases. Another interesting factor is that larger training power can trade fewer training symbols for each beam (see Fig.~\ref{fig:Np:pa:pd:1}(b)) to maintain the same misalignment probability, thus saving more symbols/time for data transmission.}

\begin{figure}[htbp]
	\centering
	\includegraphics[width=0.45\textwidth]{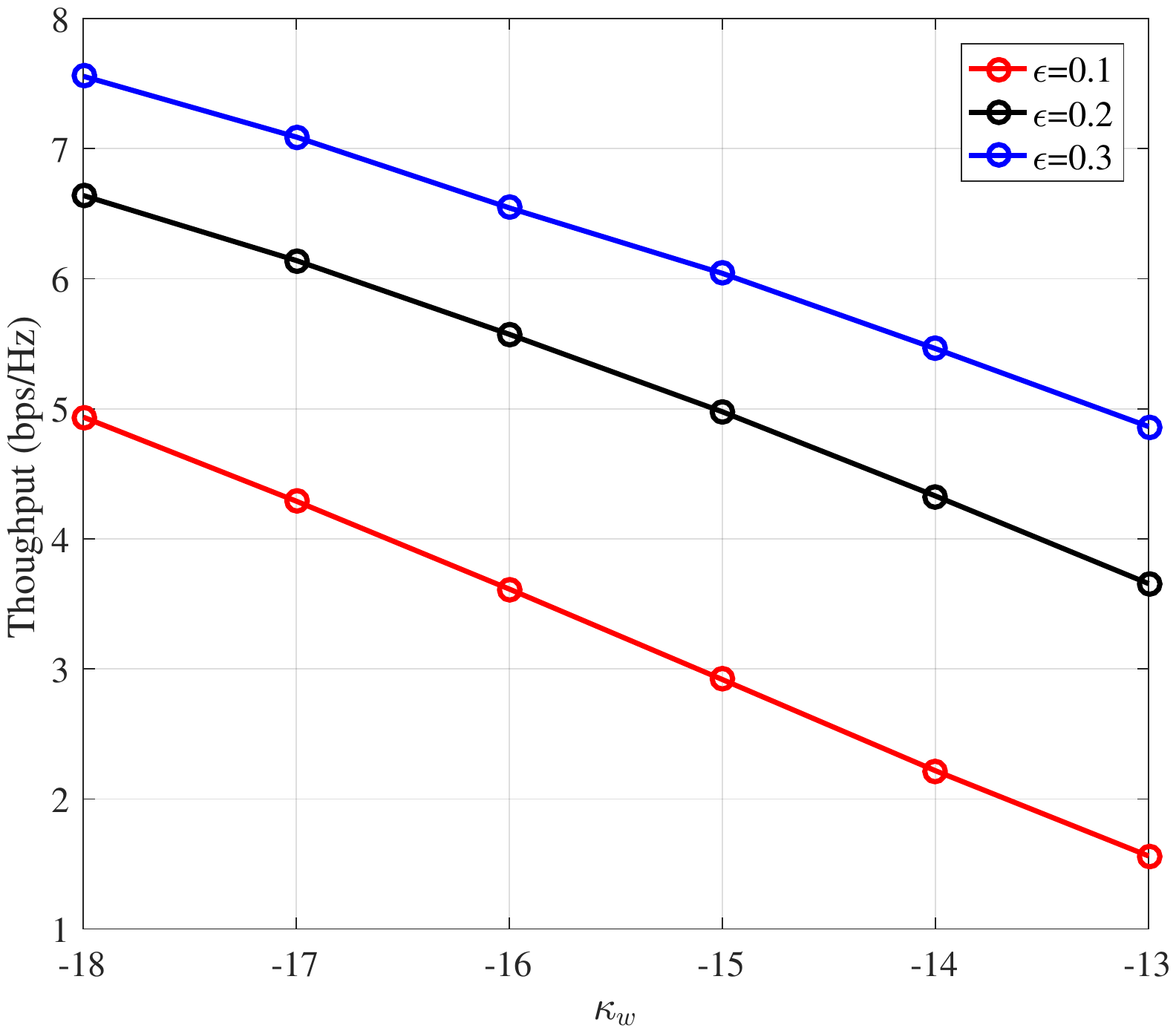}
	\caption{ Optimized $\overline{T}_{\textrm{LB}}^\star$ with different $\kappa_w$~($\kappa_b=-5\text{dB}$).}
	\label{fig:kw:vary:TLB}
\end{figure}

\begin{figure}[htbp]
	\centering
	\includegraphics[width=0.45\textwidth]{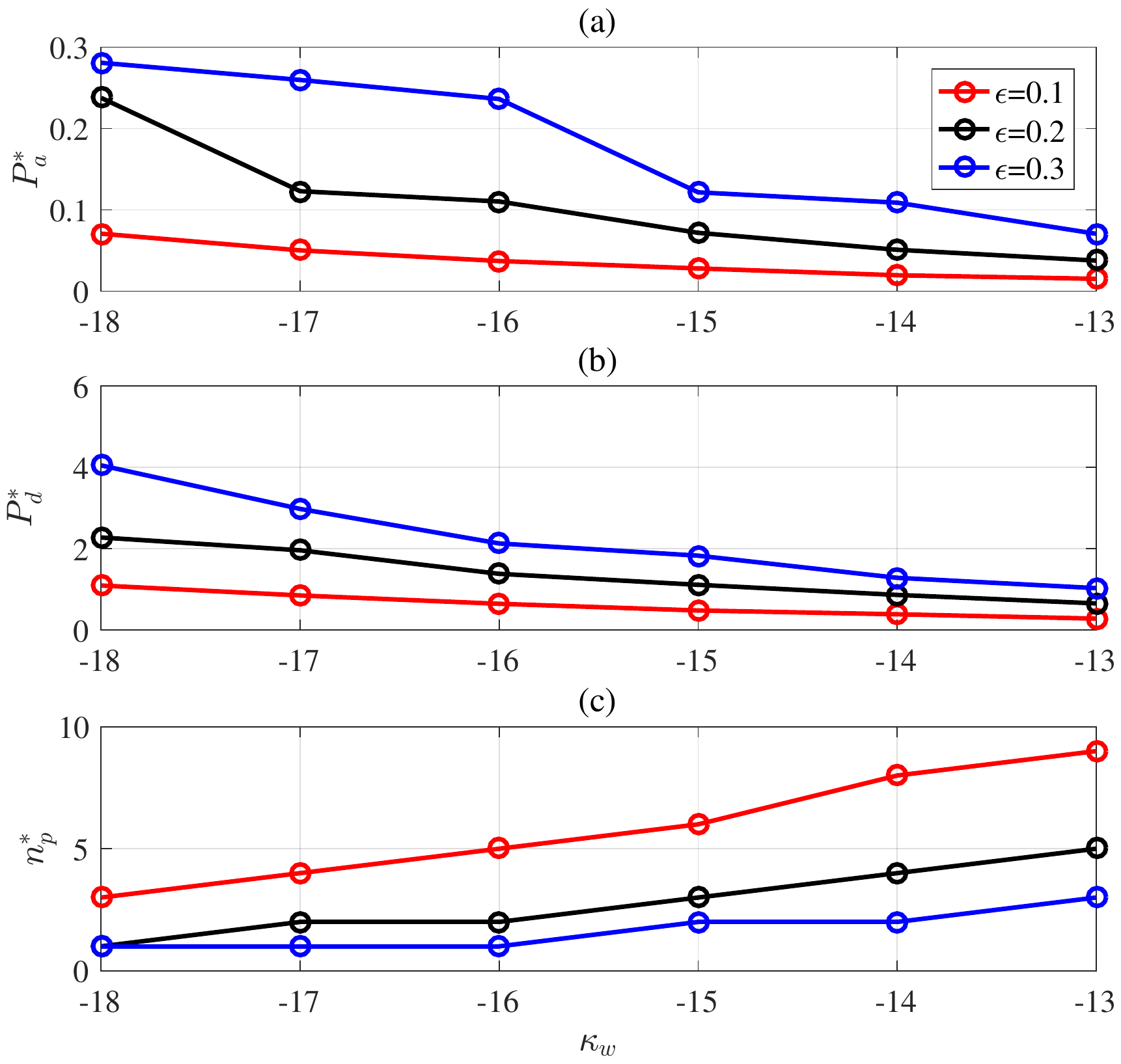}
	\caption{The optimal $(P_a^\star, P_d^\star,n_\textrm{p}^\star)$ with different $\kappa_w$~($\kappa_b=-5\text{dB}$).}
	\label{fig:kw:vary:NpP}
\end{figure}

\par With $\kappa_b = -5$ dB, Fig.~\ref{fig:kw:vary:TLB} further plots $\overline{T}_{\textrm{LB}}^\star$ versus $\kappa_w$ under three covertness levels ($\epsilon = 0.1, 0.2, 0.3$) at Willie. In all cases, Alice-Bob link throughput drops as $\kappa_w$ increases (e.g., Willie is closer to Alice). This is mainly due to the fact that training power and data transmission power at Alice have to be reduced (see Fig.~\ref{fig:kw:vary:NpP} (a)-(b)) so as to effectively hide the communication from Willie at the covertness level required. In addition, the reduction of training power in turns requires a slight increase of symbol overhead for beam training (see Fig.~\ref{fig:kw:vary:NpP} (c)) to maintain a reasonable alignment performance and effective throughput. This set of results again demonstrates the tradeoff among the key system design parameters and also the necessity of joint design of beam training and data transmission for covert mmWave communication.

\begin{figure}[htbp]
	\centering
	\includegraphics[width=0.45\textwidth]{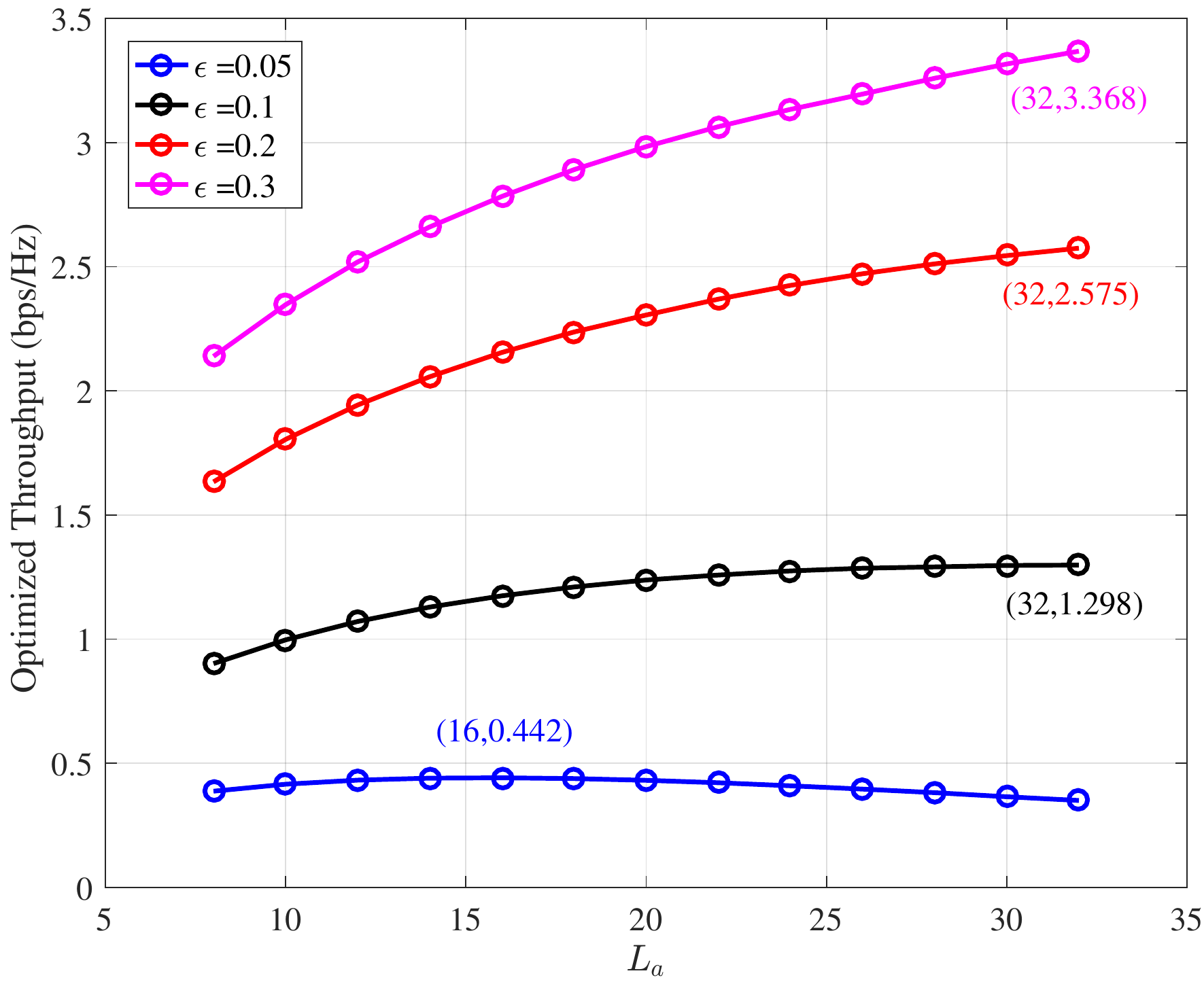}
	\caption{ Optimized $\overline{T}_{\textrm{LB}}^\star$ versus $L_a$ ($\kappa_b=-5\text{dB}, \kappa_w=-15\text{dB}$).}
	\label{fig:La:vary:TLB}
\end{figure}

\par We finally turn to investigate the impact of the main-lobe beamwith of the training codebooks on the system performance. In particular, with $L_b =8$ fixed at Bob, the number of beams at Allice $L_a$ varies from $8$ to $32$, leading to more beams with narrower main-lobe beamwith and higher beamforming gain. Having narrower beams would be beneficial for Alice-Bob link if a successful  alignment can be made. However, this might increase the total training overhead as more beams need to be examined and also increase the chance of being detected when a narrow beam trained pointing towards Willie. Fig.~\ref{fig:La:vary:TLB} depicts the optimized $\overline{T}_{\textrm{LB}}^\star$ versus $L_a$ with $\kappa_b = -5$~dB and $\kappa_w = -15$~dB. It can be seen that when the covertness requirement at Willie is not that stringent (e.g., $\epsilon = 0.1,0.2$ or $0.3$), narrow-beam codebook with $L_a = 32$ achieves the largest throughput. On the other hand, when the covertness requirement is high, the largest throughput is attained by relatively wide-beam codebook instead, e.g., codebook with $L_a = 16$ beams only for $\epsilon = 0.05$.

%

\section{Conclusions} \label{sec:conclusions}
\par In this paper, we have considered the joint design of beam training and data
transmission for a covert mmWave communication system. In particular, we have developed a framework of jointly optimizing the beam
training duration, training power and data transmission power to maximize the throughput of Alice-Bob link while ensuring the covertness constraint at Willie is met. Our analytical and numerical studies have demonstrated interesting tradeoff between the the legitimate link's throughput performance and the achievable covertness level against Willie, which thus lead to important guidelines on the design and optimization of covert mmWave communication. As future work, practical beam codebooks at the multi-antenna parties and more sophisticated channel models can be incorporated into the framework and their impacts on the performance of such a covert mmWave system can be examined. Moreover, it is also of great interest to develop some advanced covert beam training strategies that can further reduce the training overhead and boost the covert rate for the legitimate link.

\appendices
\section{{Proof of Proposition~\ref{prop:align}}} \label{appendix:prop:palign}
\par Recalling from~\eqref{equ:p:align}, $p_{\text{align}}$ is represented as:
\begin{align}
p_{\text{align}} = 1 - \Pr\{T_1\le \max\{T_2, T_3,\cdots, T_L\}\},
\end{align}
where $\Pr\{T_1\le \max\{T_2, T_3,\cdots, T_L\}\}$ corresponds to the probability of misalignment (denoted by $p_{\text{miss}}$) for Alice-Bob link. To establish a lower bound on $p_{\text{align}}$, we can derive an upper bound on $p_{\text{miss}}$. In particular, we note that
\begin{align}
&p_{\text{miss}}= \Pr\{T_1\le \max\{T_2, T_3,\cdots, T_L\}\}  \nonumber \\
&= \Pr \left\{ \begin{array}{c}
T_1  \le \max \{ T_2 ,T_3 , \cdots ,T_{L_a  + 1} \}, \\
 \text{or}~T_1  \le \max  \{ T_{L_a  + 2} ,T_{L_a  + 3} , \cdots ,T_{L_a  + L_b  + 1}\},\\
 \text{or}~T_1  \le \max \{ T_{L_a  + L_b  + 2} ,T_{L_a  + L_b  + 3} , \cdots ,T_L \} \\
 \end{array} \right\} \nonumber \\
&\le \Pr \left\{ T_1  \le \max \{ T_2 ,T_3 , \cdots ,T_{L_a  + 1} \}\right\} \nonumber \\
  &~~+\Pr \left\{ T_1  \le \max  \{ T_{L_a  + 2} ,T_{L_a  + 3} , \cdots ,T_{L_a  + L_b  + 1}\}\right\} \nonumber \\
  &~~+\Pr \left\{T_1  \le \max \{ T_{L_a  + L_b  + 2} ,T_{L_a  + L_b  + 3} , \cdots ,T_L \}\right\} \label{equ:appendix:union:bound}\\
& \triangleq {p}_{\text{miss},1} + {p}_{\text{miss},2} +  {p}_{\text{miss},3}, \nonumber
\end{align}
where~\eqref{equ:appendix:union:bound} follows from the union bound. Moreover, by Lemma~\ref{lemma:statistics} and the discussions thereafter, we have that variable $T_1 \sim \chi _2^2 \left(\lambda_A\right)$, while variables $\{T_2,\cdots, T_{L_a+1}\} \sim \chi _2^2 \left(\lambda_B\right)$, $\{T_{L_a+2},\cdots, T_{L_a+L_b+1}\} \sim \chi _2^2 \left(\lambda_C\right)$ and $\{T_{L_a+L_b+2},\cdots, T_{L}\} \sim \chi _2^2 \left(\lambda_D\right)$. Therefore, ${p}_{\text{miss},1}$ can be evaluated as
\begin{align}
{p}_{\text{miss},1} &= \Pr \left\{ T_1  \le \max \{ T_2 ,T_3 , \cdots ,T_{L_a  + 1} \}\right\} \nonumber\\
& = 1-\Pr \left\{ T_1  > \max \{ T_2 ,T_3 , \cdots ,T_{L_a  + 1} \}\right\} \\
& = 1-\Pr \left\{ T_2 < T_1, T_3 < T_1, \cdots, T_{L_a  + 1} < T_1\right\} \\
& = 1-\int_{\rm{0}}^\infty  {\left( {F\left( {t\left| {2,\lambda _B } \right.} \right)} \right)^{L_a  - 1} f\left( {t\left| {2,\lambda _A } \right.} \right)dt},
\end{align}
where $f\left( {t\left| {k,\lambda} \right.} \right)$ and $F\left( {t\left| {k,\lambda} \right.} \right)$ are the pdf and cdf of $\chi _k^2 \left( \lambda  \right)$, respectively. The other two terms ${p}_{\text{miss},2}$ and ${p}_{\text{miss},3}$ can be evaluated similarly. In this way, we establish an upper bound on $p_{\text{miss}}$ as $p_{\text{miss}} \le p_{\text{miss},1}+ p_{\text{miss},2} + p_{\text{miss},3}$, which leads to a lower bound on $p_{\text{align}}$ as $p_{\text{align}} \ge {p}_{\text{LB}} =  1-p_{\text{miss},1} - p_{\text{miss},2} - p_{\text{miss},3}$.

{
\section{{Proof of Proposition~\ref{tho:Convergence}}} \label{appendix:prop:cov}
	We first prove the existence of at least one limiting point before stating that any limiting point generated by Algorithm \ref{alg:DSCA} is a stationary solution. In this paper, the feasible set of each variable block is compact, respectively. In addition, the problem  \eqref{equ:optm:o} over the feasible region is bounded. Hence, the sequence of iterates $P^t_a$, $n^t_p$, $P^t_d$ generated by Algorithm  \ref{alg:DSCA} is compact and bounded. Since any compact and bounded sequence must have at least one limiting point, the existence of a limiting point of Algorithm \ref{alg:DSCA} is guaranteed.
	
	Let $\bm{\theta}^{\star}\triangleq (P_a^\star,P_d^\star,n_p^\star) $ and $\nu^{\star}$ be the primal and dual optimal points generated by Algorithm \ref{alg:DSCA}. For convenience, the KKT conditions for problem \eqref{eq:transformedP} are elaborated below:
\begin{align}
\bigtriangledown f(\bm{\theta^\star}) + \nu^{\star}\left( \mathcal D^t({\mathbb P}_0||\overline{{\mathbb P}}_1)(\bm{\theta}^{\star})-2\epsilon^2
\right)&=0,\nonumber\\
\nu^{\star}\left(\mathcal D^t({\mathbb P}_0||\overline{{\mathbb P}}_1)(\bm{\theta}^{\star})-2\epsilon^2\right)&= 0,\nonumber\\
\mathcal D^t({\mathbb P}_0||\overline{{\mathbb P}}_1)(\bm{\theta}^{\star})-2\epsilon^2&\leq 0,~\nu^{\star}\geq 0.\nonumber
\end{align}
All these conditions follows immediately from the fact that  the surrogate functions designed in the  Algorithm  \ref{alg:DSCA} satisfy the function value consistency, gradient consistency and lower bound conditions defined in \cite{SCA}, which can be easily verified. Consequently, we can conclude that the objective function value is nondecreasing after each iteration. Therefore, the limiting point $\bm{\theta}^{\star}$  is a KKT point of problem  \eqref{eq:transformedP}.
Note that \eqref{equ:optm:o} and \eqref{eq:transformedP} are equivalent, in the sense that the optimal solution $\bm{\theta}^{\star}$ for the two problems are identical. This concludes that the limiting point produced by Algorithm  \ref{alg:DSCA} is a KKT point of problem \eqref{equ:optm:o}. This completes the proof.
}

\bibliographystyle{IEEETran}
\bibliography{IEEEabrv,RefBeamAlignment}

\end{document}